\documentclass[10pt,twocolumn,letterpaper]{article}

\usepackage{iccv}
\usepackage{times}
\usepackage{epsfig}
\usepackage{graphicx}
\usepackage{amsmath}
\usepackage{amssymb}
\usepackage[table,xcdraw]{xcolor}

\usepackage{booktabs} 
\usepackage{multirow} 
\usepackage{makecell} 
\usepackage{colortbl}
\usepackage{algpseudocode}
\usepackage{caption} 
\usepackage{float} 
\usepackage{tcolorbox} 
\usepackage{fvextra} 
\tcbuselibrary{breakable} 
\usepackage{tabularx}
\usepackage[ruled,vlined]{algorithm2e}

\usepackage[breaklinks=true,bookmarks=false]{hyperref}

\iccvfinalcopy 

\def\iccvPaperID{****} 

\ificcvfinal\fi

\begin{document}

\title{LFRAG: Layout-oriented Fine-grained Retrieval-Augmented Generation on Multimodal Document Understanding}


\author{Yifan Zhu\\
Zhejiang University\\
Hangzhou, China\\
{\tt\small zhuyifan25@zju.edu.cn}\\
\and
Yu Mi\\
Zhejiang University\\
Hangzhou, China\\
{\tt\small miyu@mail.sdu.edu.cn}\\
\and
Yue Lu\\
Zhejiang University\\
Hangzhou, China\\
{\tt\small yuelu\_@hotmail.com}\\
\and
Yanchu Guan\\
Hangzhou High-Tech Zone (Binjiang) \\
Institute of Blockchain and Data Security\\
Hangzhou, China\\
\and
Zhixuan Chu*\\
Zhejiang University\\
Hangzhou, China\\
{\tt\small chuzhixuan94@gmail.com}
}

\maketitle
\ificcvfinal\thispagestyle{empty}\fi

\begin{abstract}
Multimodal Retrieval-Augmented Generation (RAG) has emerged as an effective paradigm for enhancing Large Language Models (LLMs) with external knowledge. 
However, existing multimodal RAG systems predominantly rely on coarse-grained page-level retrieval, which fails to capture fine-grained semantic and layout structures in visually rich documents, thereby compromising retrieval accuracy and leading to redundant context in downstream tasks.
To address these issues, we propose Layout-oriented Fine-grained Retrieval-Augmented Generation (LFRAG), a novel framework that advances multimodal RAG from page-level to block-level retrieval. 
We perform layout segmentation to construct semantically coherent fine-grained retrieval units and design a semantic–layout fusion encoder that integrates local semantics with global context via cross-attention. 
With block-level late interaction retrieval, LFRAG enables precise query–content alignment and reduces irrelevant content for downstream generation.
To enable rigorous evaluation, we construct LFDocQA, a large-scale benchmark with block-level annotations spanning diverse document types, designed to assess both multimodal document retrieval and question answering with greater granularity than existing datasets.
Extensive experiments on LFDocQA demonstrate that LFRAG achieves state-of-the-art performance on retrieval tasks, outperforms the best baseline by 7.20\% in answer accuracy, and reduces token consumption by 73.07\% in generation tasks, confirming LFRAG as an accurate and efficient framework for multimodal RAG over visually rich documents.
Our code and datasets will be released soon.
\end{abstract}

\begin{figure}[!t]
  \centering
  \includegraphics[width=\linewidth]{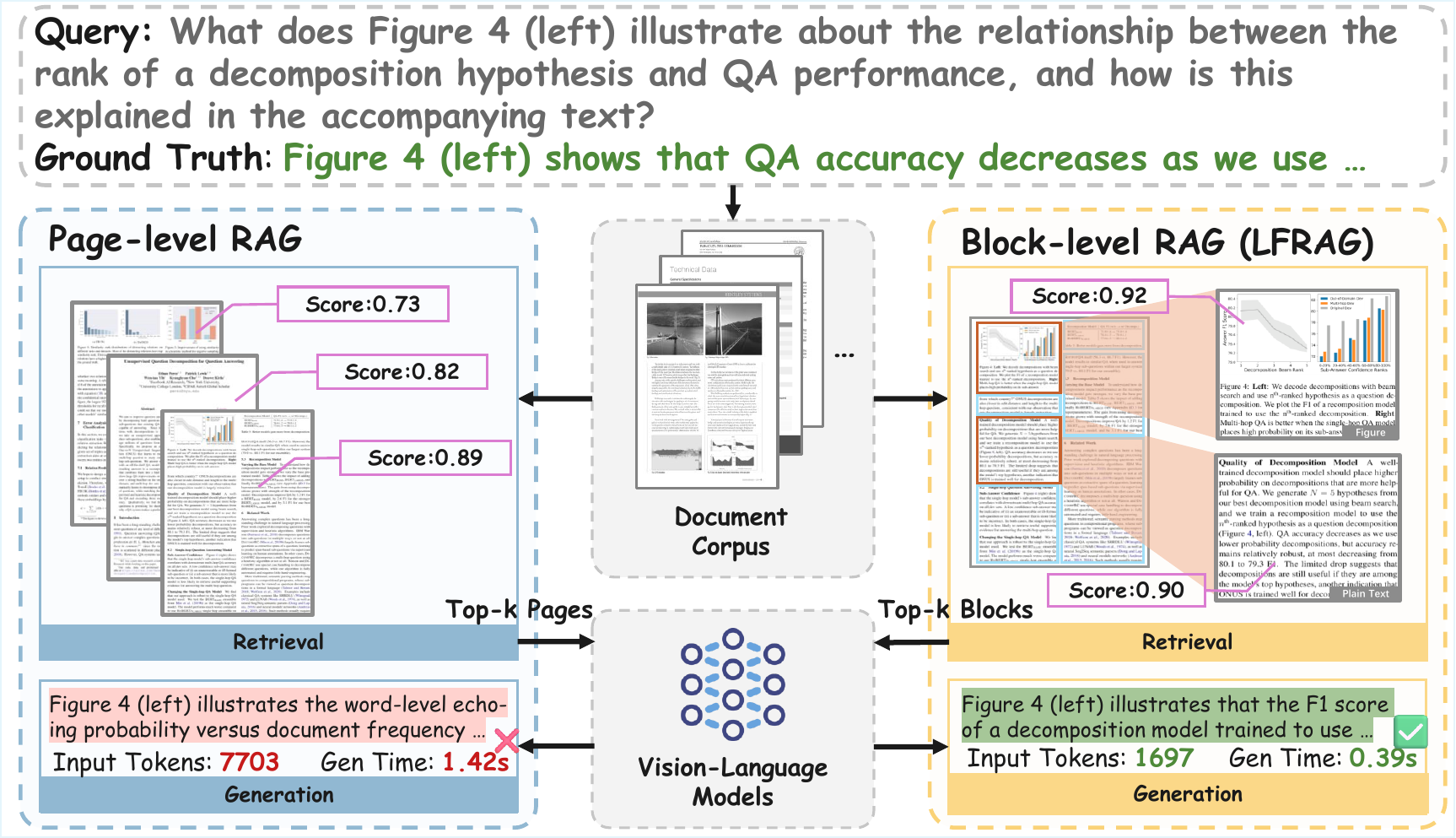}
  \caption{Comparison between page-level and layout-oriented block-level retrieval.
Our LFRAG significantly improves retrieval accuracy and generation efficiency by reducing irrelevant context and token consumption compared to page-level RAG.}
\label{figure:example}
\end{figure}

\section{Introduction}
\label{sec:intro}

Retrieval-Augmented Generation (RAG) empowers Large Language Models (LLMs) with access to external knowledge resources \cite{gao2023retrieval}. While significant progress has been made in text-based retrieval \cite{huang2024survey}, a wealth of real-world knowledge resides in visually rich documents such as scientific papers, financial reports, and infographics, whose intricate layout structures and multimodal content cannot be fully represented by plain text alone. Effectively capturing and fusing semantic content with visual structural information thus makes multimodal RAG an active research frontier at the intersection of NLP and CV.

Existing multimodal RAG methods fall into two categories. OCR-based methods \cite{BM25,leenv, chen2024bge} convert documents to plain text before retrieval, inevitably discarding visual and structural information such as table row-column relationships, chart trends, and figure-text layout logic. VLM-based methods, represented by ColPali \cite{fayssecolpali}, encode page images directly into multi-vector embeddings and retrieve by vector similarity. However, they introduce three key limitations: (1) page-level retrieval treats an entire page as the minimal retrieval unit, ignoring intra-page hierarchical semantics; (2) fixed-resolution image patching forces high-resolution pages to be downsampled, causing fine-grained information loss; and (3) the absence of layout-oriented modeling leads to poor performance on complex-layout and structured tabular content.


The root cause of these limitations is page-level retrieval granularity. Returning entire pages inevitably introduces substantial irrelevant content, increasing the computational overhead of downstream generation and introducing distracting context that elevates the risk of hallucination. Precise localization of query-relevant content is therefore central to reliable multimodal RAG.

A natural solution is to shift the minimal retrieval unit from entire pages to semantically coherent document blocks. To this end, we propose block-level retrieval, which treats each such block as an independent retrieval unit, enabling precise localization of query-relevant content, filtering out redundant context, and fundamentally improving retrieval accuracy and mitigating hallucination in generation. Nevertheless, block-level retrieval for multimodal documents faces two key challenges: how to segment documents into semantically complete blocks while preserving layout information, and how to design a VLM-based encoding and retrieval framework that effectively integrates fine-grained block features, global page context, and inter-block spatial relationships.

To tackle these challenges, we propose Layout-oriented Fine-grained Retrieval-Augmented Generation (LFRAG), a novel multimodal RAG framework that advances the retrieval paradigm from the page-level to the block-level. The core idea of LFRAG is to integrate document layout analysis into the retrieval pipeline. First, a dedicated document layout model is used to perform initial page segmentation. To mitigate the issue of excessive fragmentation, we design a semantic block aggregation strategy that merges fine-grained regions into semantically coherent blocks. Second, we design a semantic-layout feature fusion encoder that jointly encodes block-level semantic features and global page layout context, incorporating block-type tags to enhance structural awareness and enable layout-oriented representation learning. Finally, a late interaction mechanism is employed for fine-grained block-level retrieval, and the retrieved top-$k$ relevant blocks are fed into a VLM for augmented generation. This design equips LFRAG with both fine-grained information mining capabilities and holistic layout understanding capabilities, enabling precise matching between user queries and multimodal document content. Meanwhile, it significantly reduces the irrelevant context passed to the generation model, thereby achieving more accurate evidence localization and answer generation. 

To enable rigorous fine-grained evaluation, we construct LFDocQA, the first multimodal document benchmark with block-level bounding box annotations for retrieval and QA beyond page-level metrics to precisely measure retrieval quality. Experiments show that LFRAG consistently outperforms page-level baselines in both retrieval accuracy and generation quality, and remains competitive on two public page-level benchmarks when block-level results are fused to the page-level, demonstrating strong generalization.

Our contributions are summarized as follows: 
\begin{itemize}
\item We design LFRAG, a novel multimodal RAG framework integrating layout segmentation, semantic-layout fusion encoding and block-level late interaction to enable precise retrieval and efficient generation.
\item We construct LFDocQA, the first multimodal document block-level retrieval and question answering benchmark, filling the gap of existing page-level evaluation and supporting rigorous fine-grained multimodal retrieval and QA assessment.
\item Extensive experiments demonstrate that LFRAG achieves state-of-the-art retrieval and generation performance, significantly reduces generation overhead, and exhibits strong generalization across diverse document types and evaluation settings.
\end{itemize}

\begin{figure*}[!t]
  \centering
  \includegraphics[width=\linewidth]{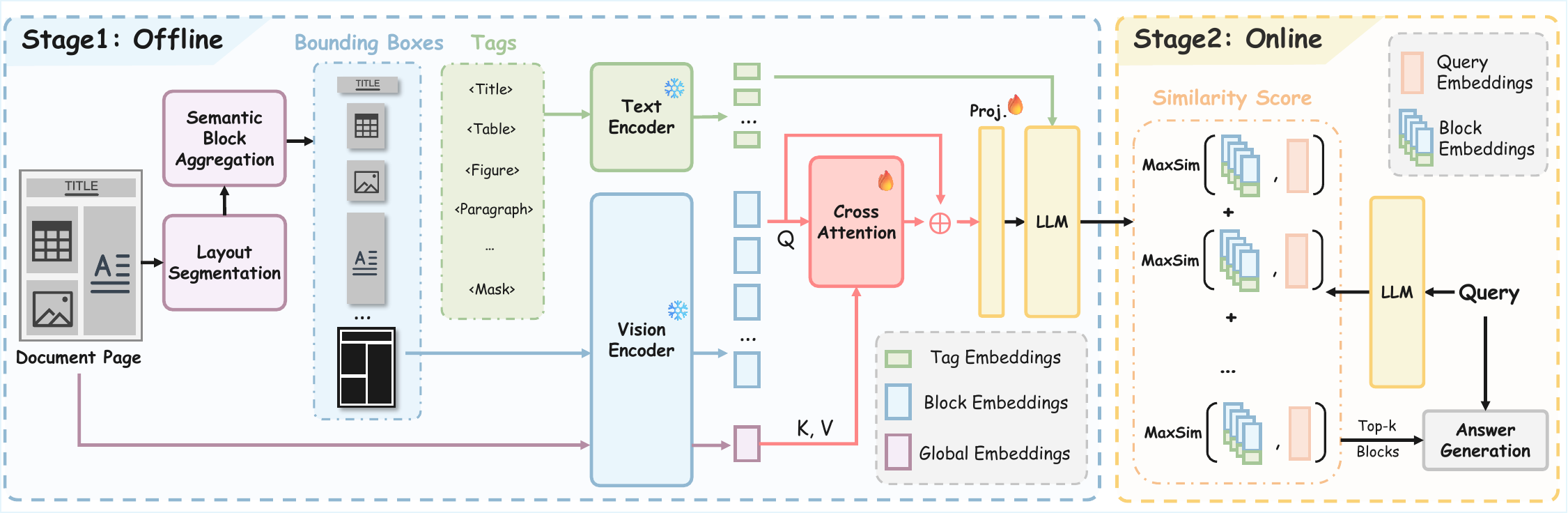}
  \caption{The framework of LFRAG. Given a query, LFRAG retrieves relevant information from a document corpus by segmenting pages into structured semantic blocks and leveraging layout-oriented representation learning for fine-grained matching, enabling more accurate evidence localization for downstream generation.}
\label{figure:pipeline}
\end{figure*}

\section{Related Works}
\subsection{Multimodal Retrieval-Augmented Generation}
Retrieval-Augmented Generation (RAG) enhances large language models (LLMs) by retrieving relevant external knowledge to ground generation and improve factuality \cite{guu2020retrieval,lewis2020retrieval,edge2024local,gao2023retrieval,huang2024survey}. 
With the increasing prevalence of visually rich documents, retrieval has extended from pure text to multimodal settings.
Early multimodal RAG typically adopt OCR-based pipelines, which convert document pages into plain text before applying text retrieval methods such as BM25 \cite{BM25}, dense retrievers like NV-Embed-V2 \cite{leenv}, or unified embedding models such as BGE-M3 \cite{chen2024bge}. 
However, these approaches inevitably lose rich visual information, especially for tables, figures, and complex layouts.
With the advancement of vision-language models (VLMs), recent research \cite{ma2024unifying,fayssecolpali,yuvisrag} shift toward directly encoding document images for retrieval. 
Methods such as ColPali \cite{fayssecolpali} employ VLMs to encode document pages as visual tokens and perform late interaction retrieval, enabling alignment between queries and image patches. Some works \cite{yuvisrag,tanaka2025vdocrag,suri2025visdom,wang2025vidorag} further integrates VLM-based retrieval into end-to-end RAG pipelines.
Despite their success, most existing methods operate at the coarse page level, retrieving entire pages as atomic units, which introduces substantial irrelevant content \cite{fayssecolpali,mace2025vidore}. 
Recent works \cite{ueda2025scan,li2026regionrag} begin to explore region-level retrieval by decomposing pages into smaller units. However, such approaches typically rely on uniform partitioning, which overlooks the intrinsic layout structure of documents and the semantic relationships between different regions.

\subsection{Retrieval-Augmented Generation Benchmarks}
Text-based RAG has long been supported by established benchmarks \cite{yang2024crag-textrag1, lyu2025crud-textrag2,chen2024benchmarking-textrag3, sorodoc2025garage-textrag4}. With the rapid development of multimodal RAG, recent benchmarks extend to visually rich documents. 
Most existing multimodal RAG benchmarks are extended from document visual question answering (DocVQA) datasets \cite{ChartQA, ArxivQA, InfoVQA, DocVQA, SlideVQA}, which provide question–answer (QA) supervision grounded in document images, forming the basis for multimodal retrieval and reasoning tasks. Building on this foundation, several benchmarks \cite{fayssecolpali, suri2025visdom, wang2025vidorag} explicitly incorporate retrieval into the evaluation pipeline. ViDoRe \cite{fayssecolpali} and VisDoMBench \cite{suri2025visdom} evaluate retrieval-augmented document understanding under realistic settings, while Document Haystacks \cite{chen2025documenthaystacks} and ViDoSeek \cite{wang2025vidorag} scale the retrieval space to thousands of documents, emphasizing large-scale retrieval and multi-document reasoning. OpenDocVQA \cite{tanaka2025vdocrag} further unifies diverse document QA datasets into an open-domain setting that requires both retrieval and reasoning across heterogeneous document sources. 
Despite these advances, existing multimodal RAG benchmarks predominantly operate at the page level, where entire document pages are treated as the basic retrieval units. This coarse-grained formulation introduces substantial irrelevant context and fails to capture fine-grained correspondences between queries and localized document regions. Moreover, the lack of block-level supervision prevents precise evaluation of retrieval quality. These limitations motivate the need for block-level benchmarks.

\begin{table*}[t]
\centering
\footnotesize
\caption{Statistics of datasets included in LFDocQA.}
\setlength{\tabcolsep}{3pt}
\renewcommand{\arraystretch}{1.0}
\begin{tabular}{l l p{2.5cm} c c c c c c}
\toprule
\textbf{Dataset} & \textbf{Domain} & \textbf{Content Type} & \textbf{QA Pairs} & \textbf{Pages} & \textbf{\makecell[c]{Avg.\\ Question Length}} & \textbf{\makecell[c]{Avg.\\ Answer Length}} & \textbf{\makecell[c]{Avg. Blocks\\ per Page}} & \textbf{\makecell[c]{Avg. Relevant Blocks\\ per Page}} \\
\midrule
Docmatix & General Documents & Text, Tables, Charts & 500 & 474 & $14.68 \pm 4.38$ & $21.49 \pm 27.44$ & $2.52 \pm 2.45$ & $1.25 \pm 0.52$ \\
PaperTab & Scientific Papers & Text, Tables, Charts & 500 & 490 & $23.32 \pm 4.70$ & $56.79 \pm 17.61$ & $3.07 \pm 1.33$ & $1.42 \pm 0.57$ \\
\midrule
LFDocQA & Combined & Text, Tables, Charts & 1000 & 964 & $19.00 \pm 6.27$ & $39.14 \pm 29.03$ & $2.79 \pm 1.99$ & $1.33 \pm 0.55$ \\
\bottomrule
\end{tabular}
\label{tab:bench_stats}
\end{table*}

\section{Methodology}
\subsection{Preliminary}

We consider the task of multimodal retrieval-augmented generation on visually rich documents. 
We define the document corpus as $\mathcal{C} = \{D_1, \dots, D_N\}$, where each document $D_i$ consists of a sequence of consecutive page images, i.e., $D_i = \{P_{i1}, P_{i2}, \dots, P_{iK}\}$. 
Each page image $P \in \mathbb{R}^{H \times W \times 3}$, where $H$ and $W$ denote the height and width of the page, and $3$ represents the RGB channels. For any page image $P$, we obtain $n$ semantically coherent document blocks after layout segmentation and block merging, denoted as $\mathcal{B}(P) = \{B_1, B_2, \dots, B_n\}$. 

Given a textual query $q$ from the user, our goal is to perform fine-grained block-level retrieval. 
Specifically, we train a retrieval model $M_R$ to retrieve the top-$k$ document blocks that are most semantically relevant to $q$ from the global block set
\begin{equation}
\mathcal{B}_{\text{all}} = \bigcup_{i=1}^{N} \ \bigcup_{P \in D_i} \mathcal{B}(P).
\end{equation}
The retrieved relevant blocks are denoted as $\mathcal{B}_q^{*} = \{B_1^{*}, B_2^{*}, \dots, B_k^{*}\} $,
where $k$ is the number of retrieved blocks.
Subsequently, the query $q$ and the retrieved relevant blocks $\mathcal{B}_q^{*}$ are fed into a generation model $M_G$ to produce an accurate answer $a$ aligned with the query.

\subsection{Layout-oriented Fine-grained Retrieval-Augmented Generation}

As illustrated in Figure \ref{figure:pipeline}, the LFRAG framework consists of four key stages: document layout segmentation and semantic block aggregation,  semantic-layout feature fusion encoding, late interaction block-level retrieval, and augmented generation, forming an end-to-end pipeline for the multimodal RAG task.

\begin{figure}[t]
  \centering
  \includegraphics[width=\linewidth]{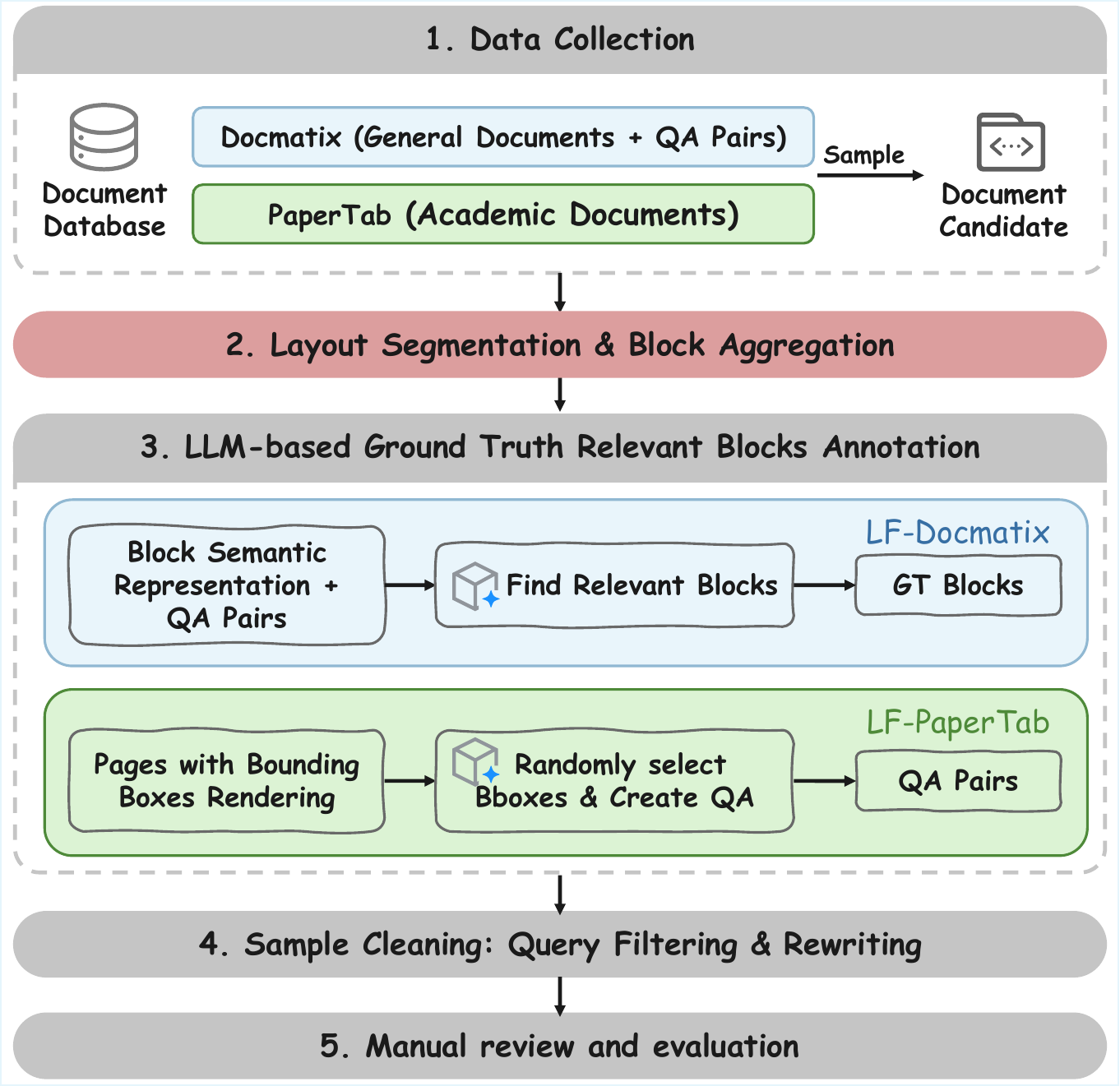}
  \caption{Construction of LFDocQA Benchmark. 
LFDocQA consists of two subsets, LF-Docmatix and LF-PaperTab, built via a semi-automatic pipeline to provide fine-grained annotations for retrieval and QA.}
\label{figure:benchmark}
\end{figure}

\noindent \textbf{Document Layout Segmentation and Semantic Block Aggregation.}
We first convert each document page into layout-oriented semantic blocks. 
Given a page image $P$, we apply a layout detection and segmentation model to identify structural regions and their corresponding labels. 
Specifically, we adopt \textit{DocLayout-YOLO} \cite{zhao2024doclayout} to generate a set of detected regions $\mathcal{R} = \{r_1, r_2, \dots, r_m\}$, where each initial region $r_i$ corresponds to a cropped image $I_i$, a bounding box $\mathrm{bbox}_i$, and a layout type tag $t_i$.
However, segmentation models often produce overly fine-grained fragments, such as separating a figure from its caption or splitting a paragraph into multiple boxes, which disrupts semantic coherence.

To address this issue, we introduce a graph-based semantic block aggregation algorithm (Algorithm~\ref{alg1}) to merge spatially adjacent and semantically compatible regions. 
Given the initial region set $\mathcal{R}$, we construct an undirected graph $G = (V, E)$ where each node corresponds to a region $r_i$. 
For each pair of regions $(r_i, r_j)$, we compute their horizontal overlap $\text{IoU}_x(r_i, r_j)$, vertical distance $\Delta y(r_i, r_j)$, and overlap ratio $\text{Ov}(r_i, r_j)$. 
We define a semantic consistency indicator $S_{ij}$ to indicate whether two regions share the same type, and a spatial compatibility condition $G_{ij}$ based on horizontal overlap and vertical alignment. 
Two regions are connected when both $S_{ij}$ and $G_{ij}$ are satisfied, or when their spatial overlap $\text{Ov}(r_i, r_j)$ exceeds a threshold $ \tau_o$. 
This formulation captures both intra-type continuity (e.g., paragraph fragments or table cells) and inter-type associations (e.g., figure–caption pairs).
The final semantic blocks are obtained as the connected components $\{C_k\}$ of graph $G$. 
For each component $C_k$, we construct a unified bounding box $\mathrm{bbox}_k$ that encloses all regions in $C_k$, and determine its type tag $t_k$ via a priority mapping. 
Each component thus forms an aggregated block $B_k$, producing refined semantic blocks $\mathcal{B}(P) = \{B_1, \dots, B_n\}$ that serve as the fundamental retrieval units in our framework.

Since cropping blocks may discard contextual information outside their bounding boxes, we further construct an additional masked page image $P^{\text{mask}}$, where all detected block regions are masked out and only the remaining background context is preserved. 
This masked page is treated as an auxiliary block $B_{n+1}$ and included in $\mathcal{B}(P)$, enabling the model to retain complementary global information that may not be fully captured within individual block crops.

\noindent \textbf{Semantic-Layout Feature Fusion Encoding.}
After obtaining layout-oriented blocks, we design a hierarchical encoding strategy that first models visual structural dependencies in the image embedding space, and then aligns representations into the textual space.

Each block image $B_i$ is encoded by the visual encoder $f_v(\cdot)$ of $M_R$, producing a block-level embedding $\mathbf{h}_i$. 
To integrate global layout context into local block representations, we additionally encode the full page $P$ into a global embedding $\mathbf{h}_{global}$ and apply a cross-attention mechanism in the visual embedding space, using the set of ${\mathbf{h}_i}$ as queries and $\mathbf{h}_{global}$ as the key and value:

\begin{equation}
\text{Attention}(Q,K,V) = \text{softmax}\left(\frac{QK^T}{\sqrt{d}}\right)V,
\end{equation}

\begin{equation}
\mathbf{h}_i^{\text{ctx}} = 
\text{Attention}(\mathbf{H},\mathbf{h}_{\text{global}}, \mathbf{h}_{\text{global}}),
\end{equation}
where $\mathbf{H} = [\mathbf{h}_1, \dots, \mathbf{h}_n]^\top$. 
The contextualized visual representation for each block is obtained by concatenating its original embedding with the attended output:
\begin{equation}
\mathbf{h}_i' = \mathbf{h}_i \parallel \mathbf{h}_i^{\text{ctx}}.
\end{equation}
The fused visual embedding is then projected into the textual embedding space via a learnable projection matrix $W_p$:
\begin{equation}
\tilde{\mathbf{h}}_i = W_p \mathbf{h}_i'.
\end{equation}
Meanwhile, the layout tag $t_i$ is encoded by the text encoder $f_t(\cdot)$ of $M_R$ to generate a tag embedding $\mathbf{t}_i$. The final block representation used for retrieval is constructed by concatenating the projected visual embedding with the tag embedding $\mathbf{e}_B^{(i)} = \tilde{\mathbf{h}}_i \parallel \mathbf{t}_i$.

\begin{algorithm}[t]
\SetAlgoLined
\caption{Semantic Block Aggregation}
\KwIn{Candidate regions $\mathcal{R} = \{r_i\}_{i=1}^{m}$ (ordered by reading order); thresholds $\tau_x, \tau_y, \tau_o, \delta$}
\KwOut{Aggregated semantic blocks $\mathcal{B}$}

Construct an undirected graph $G = (V, E)$ with $V = \{1, \dots, m\}$\;

\For{$i \gets 1$ \KwTo $m$}{
    \For{$j \gets i+1$ \KwTo $m$}{
        Compute $\text{IoU}_x(r_i, r_j)$, $\Delta y(r_i, r_j)$, and $\text{Ov}(r_i, r_j)$\;
        
        $S_{ij} \gets (t_i = t_j)$\;
        $G_{ij} \gets (\text{IoU}_x(r_i, r_j) > \tau_x) \land (-\delta < \Delta y(r_i, r_j) < \tau_y)$\;
        $M_{ij} \gets (S_{ij} \land G_{ij}) \lor (\text{Ov}(r_i, r_j) > \tau_o)$\;
        
        \If{$M_{ij} = 1$}{
            Add edge $(i, j)$ to $E$\;
        }
    }
}

Compute connected components $\{C_k\}_{i=1}^{n}$ of graph $G$\;

\ForEach{$C_k$}{
    $\mathrm{bbox}_k \gets \left( \min_{i \in C_k} x_{1i}, \min_{i \in C_k} y_{1i}, \max_{i \in C_k} x_{2i}, \max_{i \in C_k} y_{2i} \right)$\;
    $t_k \gets \Pi(\{t_i \mid i \in C_k\})$\;
    $B_k \gets (\mathrm{bbox}_k, t_k)$\;
}

\Return $\mathcal{B} = \{B_k\}_{i=1}^{n}$\;
\label{alg1}
\end{algorithm}

\noindent \textbf{Block-level Late Interaction.}
For retrieval, the query $q$ is encoded by $f_t(\cdot)$ into token-level embeddings $E_q = \{\mathbf{e}_q^{(1)}, \dots, \mathbf{e}_q^{(i)}\}$.
Given a document block $B$, we denote their multi-vector 
representations in the shared embedding space $\mathbb{R}^D$ as:

\begin{equation}
\mathbf{E}_q \in \mathbb{R}^{N_q \times D}, 
\qquad
\mathbf{E}_B \in \mathbb{R}^{N_B \times D},
\end{equation}
where $N_q$ and $N_B$ denote the number of token embeddings in the query and in the block representation, respectively. Following the late interaction in ColBert \cite{khattab2020colbert}, the relevance between $q$ and $B$ is computed using the MaxSim score:
\begin{equation}
\mathrm{S}(q, B)
=
\sum_{i=1}^{N_q}
\max_{j \in [1, N_B]}
\langle
\mathbf{E}_q^{(i)}, 
\mathbf{E}_B^{(j)}
\rangle,
\label{equ-score}
\end{equation}
where $\langle \cdot , \cdot \rangle$ denotes the dot product in $\mathbb{R}^D$. 
Top-$k$ blocks are retrieved based on $S(q, B_i)$ for downstream generation.

During training, we optimize a contrastive objective.
Consider a mini-batch consisting of $b$ query–page pairs $\{(q_k, P_k)\}_{k=1}^{b}$. 
Let $\mathcal{B}^{\text{batch}}$ denote the set of all blocks extracted from all pages $\{P_k\}_{k=1}^{b}$ in the current batch:
\begin{equation}
\mathcal{B}^{\text{batch}}
=
\bigcup_{k=1}^{b}
\mathcal{B}(P_k),
\end{equation}
where $\mathcal{B}(P_k)$ is the block set obtained from page $P_k$.
For each query $q_k$, we define its positive block set 
$\mathcal{B}_{q_k}^{+} \subset \mathcal{B}(P_k)$ 
as all blocks annotated as relevant to $q_k$. 
All other blocks within the same batch are treated as negatives:
\begin{equation}
\mathcal{B}_{q_k}^{-}
=
\mathcal{B}^{\text{batch}}
\setminus
\mathcal{B}_{q_k}^{+}.
\end{equation}
Specifically, this negative set includes:
(1) non-relevant blocks from the same page as $q_k$, and  
(2) all blocks from other pages in the batch.
We adopt a multi-positive softmax contrastive objective:

\begin{footnotesize}
\begin{equation}
\mathcal{L}
=
-\frac{1}{b}
\sum_{k=1}^{b}
\log
\frac{
\sum_{B \in \mathcal{B}_{q_k}^{+}}
\exp\big(
\mathrm{S}(q_k, B)/\tau
\big)
}{
\sum_{B \in \mathcal{B}_{q_k}^{+}}
\exp\big(
\mathrm{S}(q_k, B)/\tau
\big)
+
\sum_{B \in \mathcal{B}_{q_k}^{-}}
\exp\big(
\mathrm{S}(q_k, B)/\tau
\big)
},
\end{equation}
\end{footnotesize}

where $\tau$ is a temperature hyperparameter. This objective encourages the model to assign higher similarity to relevant blocks while discriminating against irrelevant regions.

\noindent \textbf{Retrieval-Augmented Generation.}
After retrieving the top-$k$ blocks $\mathcal{B}_q^{*}$, we perform retrieval-augmented generation. 
Specifically, the query $q$ and the retrieved block images are jointly fed into a VLM to produce the final answer:
\begin{equation}
a \xleftarrow{} M_G\big(q, \mathcal{B}_q^{*}), \ \ 
\mathcal{B}_q^{*} \xleftarrow{} \operatorname{TopK}_{B \in \mathcal{B}_{\text{all}}} \ \mathrm{S}(q, B).
\end{equation}

By restricting the input context to fine-grained relevant blocks rather than entire pages, the generation model operates on structurally grounded evidence, thereby reducing irrelevant information and mitigating hallucination risks.

\begin{table*}
    \centering
    \footnotesize
    \caption{Comparison of block-level and page-level retrieval performance between our LFRAG and other baseline models on LFDocQA. We present the nDCG@3 and Recall@3 metrics on the LF-Docmatix (in-domain) and LF-PaperTab (out-of-domain) subsets, highlighting the superiority of our LFRAG in both granularity retrieval tasks.}
    \begin{tabular}{l l c c c c c c c c}
        \toprule 
        \multirow{3}{*}{\textbf{Type}} & \multirow{3}{*}{\textbf{Model}} & \multicolumn{4}{c}{\textbf{Block-level}} & \multicolumn{4}{c}{\textbf{Page-level}} \\
        & & \multicolumn{2}{c}{\textbf{LF-Docmatix}} & \multicolumn{2}{c}{\textbf{LF-PaperTab}} & \multicolumn{2}{c}{\textbf{LF-Docmatix}} & \multicolumn{2}{c}{\textbf{LF-PaperTab}} \\
        & & nDCG@3 & Recall@3 & nDCG@3 & Recall@3 & nDCG@3 & Recall@3 & nDCG@3 & Recall@3 \\
        \cmidrule(l){1-1} \cmidrule(l){2-2} \cmidrule(lr){3-6} \cmidrule(lr){7-10}
        \multirow{3}{*}{\makecell{OCR-based}} & BM25 & 58.26 & 61.31 & 73.20 & 74.46 & 76.19 & 80.40 & 84.49 & 88.80 \\
        & BGE-M3 & 63.46 & 66.71 & 66.85 & 69.30 & 77.80 & 82.00 & 64.73 & 69.20 \\
        & NV-Embed-V2 & 73.74 & 78.48 & 73.60 & 76.43 & 84.72 & 88.60 & 78.23 & 81.80 \\
        \midrule 
        \multirow{5}{*}{\makecell{VLM-based}} & SigLIP & 55.97 & 63.13 & 30.24 & 33.60 & 79.55 & 83.80 & 48.13 & 54.00 \\
        & VisRAG & 81.49 & 87.35 & 72.54 & 76.33 & 91.78 & 94.60 & 82.68 & 86.00 \\
        & ColPali-v1.1 & 78.87 & 82.53 & 82.60 & 83.73 & 87.81 & 91.00 & 86.95 & 90.00 \\
        & ColQwen2.5-v0.2 & 85.06 & 90.23 & 87.53 & 86.93 & 92.65 & 95.00 & 91.88 & 92.80 \\
        \rowcolor{yellow!20}
        & \textbf{LFRAG (Ours)} & \textbf{90.01} & \textbf{94.08} & \textbf{90.84} & \textbf{89.76} & \textbf{94.44} & \textbf{97.40} & \textbf{94.54} & \textbf{95.80} \\
        \bottomrule 
    \end{tabular}
    \label{tab:exp1-our-benchmark}
\end{table*}

\subsection{Benchmark Construction: LFDocQA}

Existing multimodal RAG benchmarks evaluate performance at the page level, where systems are required to retrieve top-$k$ relevant pages given a query. 
However, such evaluation does not reflect the fine-grained retrieval objective of LFRAG. 
To rigorously assess block-level retrieval quality, it is necessary to construct a benchmark where queries are explicitly aligned with relevant layout regions rather than entire pages.

The primary challenge in constructing such a benchmark lies in obtaining reliable bounding-box annotations that correspond to query-relevant evidence. 
Most existing document question-answer (QA) datasets provide QA pairs without explicit localization of supporting regions. 
Therefore, as shown in Figure \ref{figure:benchmark}, we design a semi-automatic data construction pipeline to build a block-level multimodal retrieval benchmark named \textbf{LFDocQA}, which consists of two subsets constructed from complementary data sources to ensure diversity in document types and query distributions.

\noindent \textbf{LF-Docmatix.} The first subset is built upon Docmatix \cite{Docmatix}, a large-scale document visual question answering dataset containing millions of images and QA pairs. 
However, it is not directly suitable for retrieval evaluation because many questions are either context-ambiguous or not retrievable in a block-level setting.
To address this limitation, we construct structured block-level annotations that support fine-grained retrieval.
Specifically, we first perform layout segmentation and semantic block aggregation on the images. For blocks labeled as \textit{figure} or \textit{table}, we use Qwen2.5-VL-7B-Instruct \cite{qwen25vl} to generate detailed visual descriptions. 
For other textual regions, we apply the OCR engine Tesseract \cite{Tesseract} to extract textual content. 
Each block is thus converted into a unified textual representation that describes its visual or textual semantics.
Given the QA pair and the set of block descriptions from the corresponding document, we employ Qwen2.5-72B-Instruct \cite{qwen25} to identify which blocks are relevant to answering the query. 
The model is prompted to select one or multiple bounding boxes whose content directly supports the answer. 
The selected blocks are treated as ground-truth relevant regions.

\noindent \textbf{LF-PaperTab.} To further enhance structural diversity and encourage retrieval over academic documents with complex layouts, we construct a second subset based on PaperTab, a sub-collection from UDA \cite{hui2024uda-wordf1}. 
PaperTab consists of academic papers containing dense tables, figures, and multi-column layouts.
For each document page, we also apply \textit{DocLayout-YOLO} to obtain bounding boxes. 
The bounding boxes are rendered directly onto the page to make spatial regions visually explicit. 
The annotated page is then provided to Qwen3-VL-32B \cite{bai2025qwen3vl}, which is prompted to randomly select several document blocks and generate corresponding QA pairs grounded in the selected regions. 
This procedure ensures that each query is explicitly tied to one or multiple bounding boxes, providing precise localization supervision.

Finally, for both subsets, we conduct query filtering and rewriting to ensure retrievability. 
We use Qwen2.5-7B-Instruct \cite{qwen25} to determine whether a given QA pair satisfies retrieval requirements. 
Questions containing ambiguous references, overly generic inquiries that admit multiple answers across documents, or meta-level document questions (e.g., ``How many paragraphs are in this document?'') are filtered out. 
Questions that are semantically valid but not explicitly retrievable will be written into structurally grounded queries that can be answered from specific document regions. All revised queries and their corresponding block annotations undergo manual review to ensure alignment with retrieval objectives and annotation accuracy.

As shown in Table \ref{tab:bench_stats}, our LFDocQA benchmark consists of two subsets. 
Each sample in LFDocQA contains: (1) a document page, (2) layout bounding boxes and structural tags, (3) a natural language query, (4) top-$k$ ground-truth relevant blocks, and (5) the ground-truth answer. 
By explicitly annotating query-block alignments, LFDocQA enables rigorous evaluation of fine-grained layout-oriented retrieval, bridging the gap between page-level retrieval benchmarks and block-level evidence localization required by retrieval-augmented generation systems. More details about datasets are provided in Appendix \ref{appendix-A-dataset}.

\section{Experiments}
\subsection{Experiment Setup}

\noindent \textbf{Training Setup.}
To train the retrieval model $M_R$, we construct a hybrid training corpus consisting of synthesized data and existing multimodal reasoning data. 
First, we build 50,787 samples following the same data construction pipeline as LF-Docmatix. 
Second, we incorporate 69,213 samples from Visual-CoT \cite{shao2024visual-cot}, a visual chain-of-thought corpus containing multimodal reasoning annotations. We retain only the document QA subset and apply the same layout segmentation and block aggregation to ensure consistent granularity. The annotated bounding boxes provided in Visual-CoT are mapped to the aggregated blocks which are regarded as positive samples.

The retrieval model $M_R$ is built upon Qwen2.5-VL-3B \cite{qwen25vl} as the backbone, augmented with an additional cross-attention layer. 
The overall training configuration follows ColPali \cite{fayssecolpali}. 
We employ low-rank adapters (LoRA) \cite{hu2022lora} on the Transformer layers of the language model and the projection layer, with rank $r=32$ and scaling factor $\alpha=32$. The newly introduced cross-attention layers are trained with full parameters. 
The model is optimized using a learning rate of $5\times10^{-5}$ with linear decay and a warm-up ratio of 2.5\%, with a batch size of 8 for 2 epochs on 8 NVIDIA H200 GPUs.
The temperature parameter in the contrastive objective is fixed to $\tau=0.02$.

\begin{table}[t]
    \centering
    \footnotesize
    \caption{Comparison of generation performance on LFDocQA. LFRAG achieves the best answer quality in terms of ROUGE-L and answer accuracy (Acc) scored by an LLM, while significantly reducing the number of input tokens.}
    \setlength{\tabcolsep}{2pt}
    \begin{tabular}{p{2.5cm} c c c c c c}
        \toprule 
        \multirow{2}{*}{\textbf{Model}} & \multicolumn{3}{c}{\textbf{LF-Docmatix}} & \multicolumn{3}{c}{\textbf{LF-PaperTab}} \\
        & rouge-L & Acc & Tokens & rouge-L & Acc & Tokens \\
        \cmidrule(l){1-1} \cmidrule(lr){2-4} \cmidrule(l){5-7}
        BM25 & 46.69 & 2.43 & 9.2k & 42.59 & 3.30 & 7.7k \\
        BGE-M3 & 49.84 & 2.46 & 8.7k & 39.22 & 3.01 & 7.6k \\
        NV-Embed-V2 & 47.85 & 2.57 & 8.8k & 40.76 & 3.22 & 7.7k \\
        SigLIP & 48.26 & 2.51 & 10.0k & 36.57 & 2.73 & 7.7k \\
        VisRAG & 46.79 & 2.62 & 8.9k & 41.68 & 3.26 & 7.8k \\
        ColPali-v1.1 & 47.88 & 2.58 & 8.9k & 42.97 & 3.34 & 7.7k \\
        ColQwen2.5-v0.2 & 47.19 & 2.65 & 9.3k & 43.09 & 3.37 & 7.7k \\
        \rowcolor{yellow!20}\textbf{LFRAG (Ours)} & \textbf{50.78} & \textbf{2.78} & \textbf{2.9k} & \textbf{45.37} & \textbf{3.69} & \textbf{1.7k} \\
        \bottomrule 
    \end{tabular}
    \label{tab:Generation-LFDocQA}
\end{table}

\begin{table*}[h]
    \centering
    \footnotesize
    \caption{Retrieval performance on the ViDoRe V1 benchmark across 10 datasets in Recall@5. LFRAG achieves the highest average score and demonstrates strong performance across multiple domains.}
    \setlength{\tabcolsep}{4pt}
    \begin{tabular}{l c c c c c c c c c c c}
        \toprule 
        \textbf{Model} & \textbf{Avg.} & \textbf{ArxivQA} & \textbf{DocVQA} & \textbf{InfoVQA} & \textbf{Shift.} & \textbf{AI} & \textbf{Energy} & \textbf{Gov. Reports} & \textbf{Healthcare} & \textbf{TabFQuad} & \textbf{TAT-DQA} \\
        \midrule 
        ColPali-v1.2 & 81.62 & 80 & 59.1 & 82.2 & 70.2 & 96.9 & 92.1 & 92.8 & 94.8 & 81.9 & 66.2 \\
        ColQwen2-v0.1 & 89.23 & \underline{88} & 61.5 & \textbf{92.5} & \underline{89.9} & \underline{99} & 95.9 & \underline{95.5} & \textbf{98.8} & \underline{89} & \underline{82.2} \\
        ColQwen2.5-v0.2 & 89.53 & \textbf{88.9} & \underline{63.6} & \textbf{92.5} & 87.9 & \textbf{99.6} & \underline{96.1} & \textbf{95.8} & \underline{98} & \textbf{90.8} & 82.1 \\
        \textbf{LFRAG (Ours)} & \textbf{90.71} & 86.5 & \textbf{66.7} & \underline{88.7} & \textbf{95.1} & \textbf{99.6} & \textbf{97.2} & \textbf{95.8} & \textbf{98.8} & 82.3 & \textbf{96.4} \\
        \bottomrule 
    \end{tabular}
    \label{tab:ViDoRe}
\end{table*}

\noindent \textbf{Evaluation Setup.}
For evaluation, we conduct experiments on LFDocQA.
We compare our LFRAG with both OCR-based pipelines and VLM-based retrieval methods. 
OCR-based pipelines include BM25 \cite{BM25}, BGE-M3 \cite{chen2024bge}, and NV-Embed-V2 \cite{leenv}, and  where document pages are first converted into text using OCR and then indexed for retrieval. 
VLM-based methods include SigLIP \cite{SigLIP}, VisRAG \cite{yuvisrag}, ColPali-v1.1 and ColQwen2.5-v0.2 \cite{fayssecolpali}. 

Retrieval performance is evaluated using NDCG@$k$ and Recall@$k$, where $k \in \{1,3,5,10\}$. 
We report results at two granularities. 
(1) \textbf{At block-level}, for LFRAG, we retrieve the top-$k$ document blocks based on the similarity score defined in Eq.~(\ref{equ-score}). 
For baselines, which are originally designed for page-level retrieval, we treat each document block as an individual page and apply their retrieval pipelines accordingly. 
(2) \textbf{At page-level}, for LFRAG, block-level similarity scores are mapped back to pages. 
Given a query $q$ and a page $P$ containing blocks $\{B_i\}$, the page-level score is defined as
\begin{equation}
S(q,P) = \max_{B_i \in P} S(q,B_i),
\label{page-level-metric}
\end{equation}
and pages are ranked based on $S(q,P)$. 
In contrast, baselines follow their original page-level retrieval pipelines.

For answer generation, we directly employ off-the-shelf generation models Qwen2-VL-2B-Instruct as $M_G$ \cite{wang2024qwen2vl} for inference without fine-tuning.
We evaluate generated answers against ground-truth using ROUGE-L \cite{lin2004rouge}. Additionally, we adopt Qwen3-14B \cite{yang2025qwen3} to scores each generated answer on a 0–5 scale via a dedicated prompt, based on its correctness and faithfulness, with details provided in Appendix \ref{appendix-B1-Setup}.

In addition, we further evaluate generalization on two public benchmarks: VisDoMBench \cite{suri2025visdom} and ViDoRe V1 \cite{fayssecolpali}, following their standard evaluation protocols. 
(1) On ViDoRe V1, we report Recall@5 for page-level retrieval. 
(2) On VisDoMBench, retrieval performance is assessed using ANLCS (Average Normalized Longest Common Subsequence) between ground truth evidence and retrieved chunks/pages. We follow the VisDomRAG pipeline: retrieved blocks are first mapped back to their corresponding pages, and then further mapped to complete documents, conducting document-level retrieval with $k$=5. 
Generation performance is evaluated using Word Overlap F1 \cite{hui2024uda-wordf1}. 


\subsection{Experimental Results on LFDocQA}


\begin{table}[t]
    \centering
    \footnotesize
    \caption{Retrieval performance of various retrieval models on 5 subsets (PaperTab, FetaTab, etc.) of VisDoMBench. LFRAG achieves the highest average ANLCS.}
    \setlength{\tabcolsep}{4pt}
    \begin{tabular}{l c c c c c c}
        \toprule 
        \textbf{Retriever} & \textbf{Paper.} & \textbf{Feta.} & \textbf{Sci.} & \textbf{SPI.} & \textbf{Slide.} & \textbf{Average} \\
        \midrule 
        BM25 & 65.51 & 84.00 & 72.73 & 88.23 & 98.55 & 81.80 \\
        MiniLM & 65.51 & 88.85 & 91.65 & 61.06 & 73.00 & 76.01 \\
        MPNet & 90.18 & 89.71 & 91.40 & 95.84 & 73.00 & 88.03 \\
        BGE1.5 & 96.81 & 94.00 & 90.91 & \underline{98.43} & 81.85 & 92.40 \\
        ColPali-v1.2 & 96.93 & 97.71 & 95.28 & 93.17 & 97.64 & 96.15 \\
        ColQwen2-v0.1 & \textbf{97.61} & 96.86 & \underline{95.58} & 96.85 & \textbf{97.82} & \underline{96.94} \\
        \textbf{LFRAG (Ours)} & \underline{97.46} & \textbf{99.71} & \textbf{98.03} & \textbf{99.49} & \underline{93.67} & \textbf{97.67} \\
        \bottomrule 
    \end{tabular}
    
    \label{tab:Visdom}
\end{table}

\noindent \textbf{Retrieval Performance.}
As shown in Table \ref{tab:exp1-our-benchmark}, we present the retrieval performance on LFDocQA (nDCG@3 and Recall@3) at both block-level and page-level across in-domain (LF-Docmatix) and out-of-domain (LF-PaperTab) settings. Complete results for nDCG@$k$ and Recall@$k$ with $k=\{1,3,5,10\}$ are provided in Appendix \ref{appendix-2-results}.

At the block level, LFRAG achieves the best performance among all baselines. 
On LF-Docmatix, LFRAG achieves nDCG@3 of 90.01 and Recall@3 of 94.08, which are 4.95 and 3.85 percentage points higher than ColQwen2.5-v0.2 (nDCG@3: 85.06, Recall@3: 90.23), which also adopts Qwen2.5-VL-3B as backbone.
Similar improvements are observed on LF-PaperTab, where LFRAG achieves nDCG@3 of 90.84 and maintains competitive recall. These results demonstrate the effectiveness of layout-oriented fine-grained retrieval in precisely locating relevant document regions.

At the page level, LFRAG continues to demonstrate its advantages. By aggregating block-level similarities (Eq.~\ref{page-level-metric}), LFRAG achieves nDCG@3 of 94.44 and Recall@3 of 97.40 on LF-Docmatix, and nDCG@3 of 94.54 and Recall@3 of 95.80 on LF-PaperTab, outperforming all baseline models by a clear margin. This indicates that fine-grained block retrieval not only improves local matching accuracy but also leads to stronger global page-level ranking.
Additionally, it is worth noting that VLM-based models generally outperform OCR-based models, indicating the importance of multimodal information in document retrieval tasks. Our LFRAG, as a VLM-based model with layout-oriented block aggregation and cross-attention mechanisms, further leverages multimodal features to achieve better retrieval performance.

\noindent \textbf{Generation Performance.}
As shown in Table \ref{tab:Generation-LFDocQA}, LFRAG achieves the highest ROUGE-L and LLM-based answer accuracy across both datasets. It indicates that fine-grained retrieval provides more relevant and precise evidence, leading to higher-quality answer generation.
Notably, LFRAG significantly reduces the number of input tokens compared to page-level retrieval methods. Specifically, it reduces token usage from around 8–10k tokens to only 2.9k on LF-Docmatix and 1.7k on LF-PaperTab, achieving 67.9\% and 77.7\% reduction in context length. 
This advantage stems from precise block-level retrieval in LFRAG, which filters out redundant information and provides only the most relevant context for generation, thus improving both efficiency and quality.

\begin{table}[t]
    \centering
    \footnotesize
    \caption{Comparison of Generation performance on VisDoMBench in Word Overlap F1.}
    \setlength{\tabcolsep}{4pt}
    \begin{tabular}{l c c c c c c}
        \toprule
        \textbf{Retriever} & \textbf{Paper.} & \textbf{Feta.} & \textbf{Sci.} & \textbf{SPI.} & \textbf{Slide.} & \textbf{Average} \\
        \midrule
        Long Context & 8.23 & 23.10 & 16.74 & 9.93 & 2.46 & 12.09 \\
        Text RAG & 25.33 & 57.56 & 26.75 & 39.77 & 8.82 & 31.65 \\
        Visual RAG & 27.37 & 58.57 & \underline{28.13} & \underline{42.81} & \underline{38.42} & 39.06 \\
        VisDoMRAG & \textbf{29.89} & \underline{59.24} & 27.98 & 42.80 & \textbf{39.77} & \underline{39.94} \\
        \textbf{LFRAG (Ours)} & \underline{28.90} & \textbf{60.71} & \textbf{29.57} & \textbf{45.33} & 36.63 & \textbf{40.23} \\
        \bottomrule
    \end{tabular}
    \label{tab:VisDom-Gen}
\end{table}

\begin{table*}[t]  
    \centering
    \caption{Ablation study on LF-Docmatix evaluating key components of LFRAG. Removing block aggregation, cross-attention, or tag embedding consistently degrades performance at both block and page levels, clarifying the contribution of each component to the overall framework.}
    \begin{tabular}{l c c c c c c c c}
        \toprule 
        \multirow{2}{*}{\textbf{Model Variant}} & \multicolumn{4}{c}{\textbf{block-level}} & \multicolumn{4}{c}{\textbf{page-level}} \\
        & N@3 & N@5 & R@3 & R@5 & N@3 & N@5 & R@3 & R@5 \\
        \cmidrule(l){1-1} \cmidrule(lr){2-5} \cmidrule(l){6-9}
        w/o Block Aggregation & 85.53 & 86.39 & 87.60 & 90.37 & 88.21 & 90.40 & 93.00 & 93.50 \\
        w/o Cross-Attention & 85.52 & 87.63 & 90.81 & 93.95 & 90.41 & 92.06 & 95.60 & 97.20 \\
        w/o Tag Embedding & 89.76 & 90.73 & 93.45 & 95.13 & 93.70 & 94.00 & 96.80 & 97.60 \\
        \rowcolor{yellow!20}
        \textbf{LFRAG} & \textbf{90.01} & \textbf{91.29} & \textbf{94.08} & \textbf{96.88} & \textbf{94.44} & \textbf{94.86} & \textbf{97.40} & \textbf{98.40} \\
        \bottomrule 
    \end{tabular}
    \label{tab:ablation1}
\end{table*}

\subsection{Generalization to Public Page-level Benchmarks}

\noindent \textbf{Results on ViDoRe V1.}
Table \ref{tab:ViDoRe} presents the retrieval performance on the ViDoRe V1 benchmark, which consists of 10 diverse datasets spanning document QA, tables, and domain-specific corpora. 
LFRAG achieves the best overall performance with an average Recall@5 of 90.71. Notably, LFRAG shows significant gains on several challenging subsets, including DocVQA (+3.1 over the best baseline) and TAT-DQA (+14.2), indicating its strong capability in handling complex and structured document understanding. 
We also observe that LFRAG underperforms slightly on a few subsets, such as TabFQuad, ArxivQA, and InfoVQA. 
For ArxivQA and InfoVQA, which are dominated by single-image inputs, the benefit of layout-oriented decomposition is limited, as the entire page often forms a single semantic unit. 
For TabFQuad, which focuses on table-centric documents in French, the performance gap is likely due to the highly structured nature of tables, where fine-grained cell-level relation-ships are critical and may not be fully captured by our block-level aggregation strategy.  In contrast, LFRAG performs strongly on ShiftProject (French), suggesting that the observed gaps are not due to language generalization but rather differences in document structure and layout characteristics.

\noindent \textbf{Results on VisDoMBench.}
We evaluate both the retrieval and the generation  performance, with detailed results presented in Tables \ref{tab:Visdom} and \ref{tab:VisDom-Gen}. LFRAG achieves the highest average ANLCS of 97.67 in retrieval tasks, outperforming ColQwen2-v0.1 by 0.73 percentage points, and ranks first in three subsets (FetaTab, SciGraphQA, SPIQA). For generation tasks, LFRAG also obtains the best average Word Overlap F1 \cite{hui2024uda-wordf1} of 40.23. We note that LFRAG shows relatively lower retrieval performance on SlideVQA compared to other subsets. This is mainly due to the unique layout characteristics of slide-style documents, where content is sparse, highly condensed, and less structured, making layout-based block decomposition less effective. Overall, the results demonstrate that LFRAG generalizes well to document-level retrieval and consistently improves downstream generation quality.

\begin{figure}[t]
  \centering
  \includegraphics[width=\linewidth]{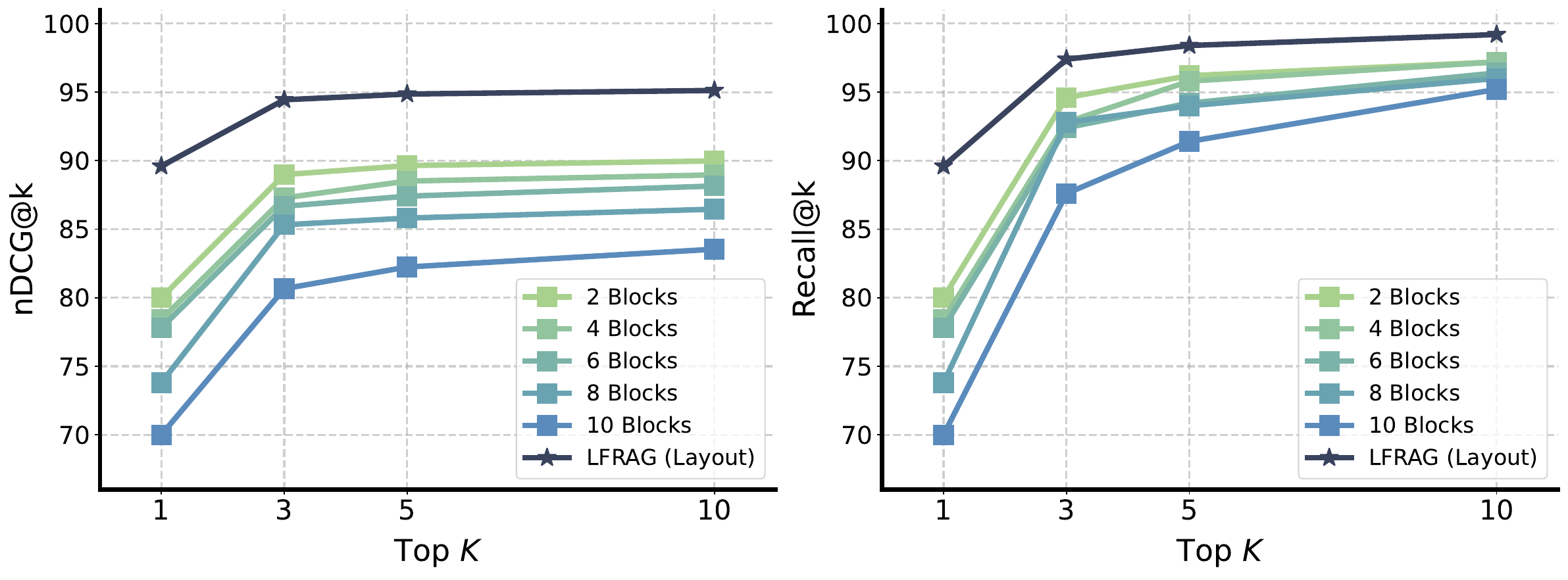}
  \caption{Comparison of page-level retrieval performance on LF-Docmatix using different segmentation strategies. Layout-oriented segmentation consistently achieves higher retrieval performance across different $k$ values.}
\label{figure:ablation2-layout}
\end{figure}

\begin{figure}[t]
  \centering
  \includegraphics[width=\linewidth]{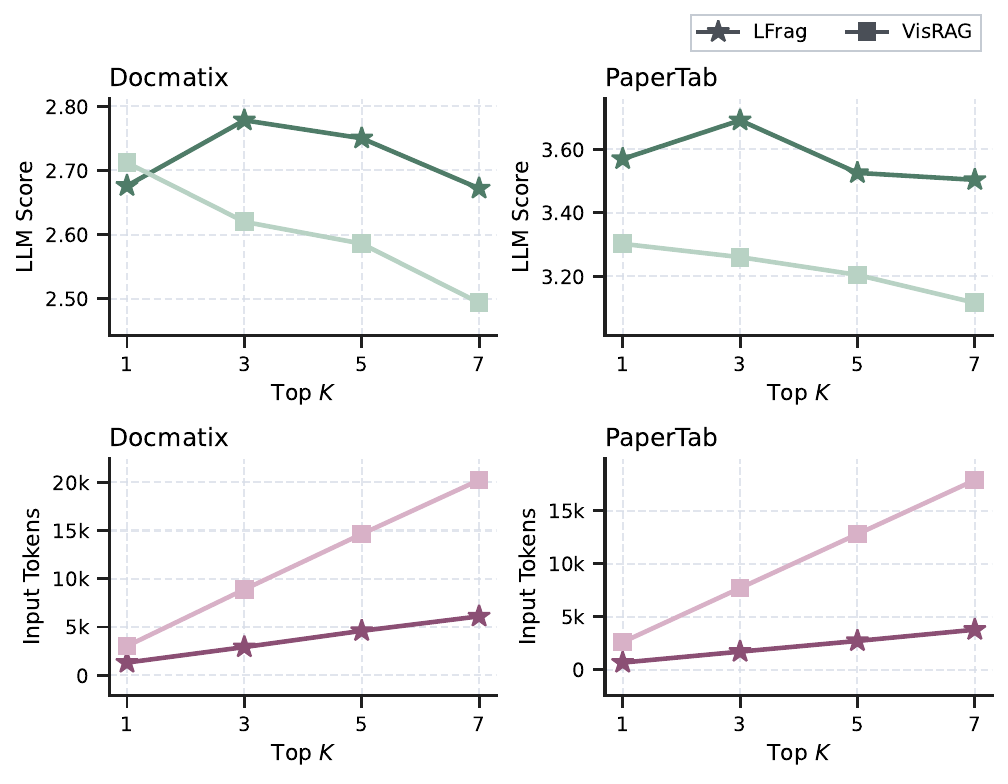}
  \caption{Impact of top-$k$ on generation performance and token usage on LFDocQA. LFRAG achieves better answer quality with significantly fewer tokens compared to page-level retrieval baselines.}
\label{figure:ablation3}
\end{figure}

\subsection{Ablation}

\noindent \textbf{Ablation on Key Components.}
Table \ref{tab:ablation1} presents the ablation study on LF-Docmatix, evaluating the contribution of key components in LFRAG, including block aggregation, cross-attention, and tag embedding. We observe consistent performance degradation across all variants at both block and page levels, confirming that each component contributes to the overall effectiveness.
In particular, removing block aggregation leads to a noticeable drop in both NDCG@$k$ (N@$k$) and Recall@$k$ (R@$k$), indicating that constructing semantically coherent units is essential for fine-grained retrieval. The absence of cross-attention also degrades performance, highlighting the importance of incorporating global context and inter-block relationships. Tag embedding contributes additional gains by injecting layout-aware semantic signals.

\noindent \textbf{Effect of Layout-oriented Segmentation.}
As shown in Figure \ref{figure:ablation2-layout}, we compare layout-oriented segmentation with fixed-size block partitioning under different numbers of blocks per page.
We find that increasing the number of uniformly partitioned blocks consistently degrades performance across all metrics. For example, NDCG@3 drops from 88.98 (2 blocks) to 80.66 (10 blocks), while Recall@3 decreases from 94.60 to 87.60. This trend indicates that naive uniform splitting disrupts semantic coherence and breaks meaningful layout structures, leading to inferior retrieval quality.
In contrast, LFRAG with layout-oriented segmentation significantly outperforms all fixed strategies, demonstrating that preserving document structure is more important than simply increasing granularity. This highlights that effective fine-grained retrieval relies on semantically meaningful segmentation rather than arbitrary splitting.

\noindent \textbf{Effect of Top-k in Generation.}
We evaluate the impact of varying $k$ on generation performance (Figure \ref{figure:ablation3}). 
LFRAG peaks at $k=3$, while larger $k$ leads to slight degradation due to redundant context. While LFRAG performs poorly at $k=1$, suggesting that a single block is often insufficient to provide complete evidence. Notably, LFRAG at $k=3$ significantly outperforms VisRAG at $k=1$, while requiring a comparable number of input tokens. This suggests that block-level retrieval provides higher information density, enabling more effective use of limited context for generation.
These results reveal a trade-off between granularity and completeness, and demonstrate that aggregating a small set of high-quality blocks yields the best balance between efficiency and answer accuracy.

\section{Conclusion}
In this paper, we introduce LFRAG, a fine-grained multimodal RAG framework that integrates layout-oriented modeling, semantic-layout fusion encoding, and block-level retrieval to enable precise document understanding. By shifting from page-level to block-level retrieval, LFRAG enables more precise alignment between queries and document content, while reducing irrelevant context for downstream generation. Our design effectively integrates semantic and structural information, leading to consistent improvements across both retrieval and generation tasks. To support rigorous evaluation, we also contribute LFDocQA, the first block-level multimodal document benchmark for retrieval and QA. In future work, we plan to extend LFRAG to better handle structured data such as tables and explore adaptive retrieval strategies that dynamically adjust granularity based on document characteristics.

{\small
\bibliographystyle{ieee_fullname}
\bibliography{main}

\begin{thebibliography}{10}\itemsep=-1pt

\bibitem{bai2025qwen3vl}
Shuai Bai, Yuxuan Cai, Ruizhe Chen, Keqin Chen, Xionghui Chen, Zesen Cheng, Lianghao Deng, Wei Ding, Chang Gao, Chunjiang Ge, et~al.
\newblock Qwen3-vl technical report.
\newblock {\em arXiv preprint arXiv:2511.21631}, 2025.

\bibitem{qwen25vl}
Shuai Bai, Keqin Chen, Xuejing Liu, Jialin Wang, Wenbin Ge, Sibo Song, Kai Dang, Peng Wang, Shijie Wang, Jun Tang, et~al.
\newblock Qwen2.5-vl technical report.
\newblock {\em ArXiv}, abs/2502.13923, 2025.

\bibitem{chen2024benchmarking-textrag3}
Jiawei Chen, Hongyu Lin, Xianpei Han, and Le Sun.
\newblock Benchmarking large language models in retrieval-augmented generation.
\newblock In {\em Proceedings of the AAAI Conference on Artificial Intelligence}, volume~38, pages 17754--17762, 2024.

\bibitem{chen2024bge}
Jianlv Chen, Shitao Xiao, Peitian Zhang, Kun Luo, Defu Lian, and Zheng Liu.
\newblock Bge m3-embedding: Multi-lingual, multi-functionality, multi-granularity text embeddings through self-knowledge distillation.
\newblock {\em arXiv preprint arXiv:2402.03216}, 4(5), 2024.

\bibitem{chen2025documenthaystacks}
Jun Chen, Dannong Xu, Junjie Fei, Chun-Mei Feng, and Mohamed Elhoseiny.
\newblock Document haystacks: Vision-language reasoning over piles of 1000+ documents.
\newblock In {\em Proceedings of the Computer Vision and Pattern Recognition Conference}, pages 24817--24826, 2025.

\bibitem{edge2024local}
Darren Edge, Ha Trinh, Newman Cheng, Joshua Bradley, Alex Chao, Apurva Mody, Steven Truitt, Dasha Metropolitansky, Robert~Osazuwa Ness, and Jonathan Larson.
\newblock From local to global: A graph rag approach to query-focused summarization.
\newblock {\em arXiv preprint arXiv:2404.16130}, 2024.

\bibitem{fayssecolpali}
Manuel Faysse, Hugues Sibille, Tony Wu, Bilel Omrani, Gautier Viaud, C{\'e}line Hudelot, and Pierre Colombo.
\newblock Colpali: Efficient document retrieval with vision language models.
\newblock {\em arXiv preprint arXiv:2407.01449}, 2024.

\bibitem{gao2023retrieval}
Yunfan Gao, Yun Xiong, Xinyu Gao, Kangxiang Jia, Jinliu Pan, Yuxi Bi, Yixin Dai, Jiawei Sun, Haofen Wang, Haofen Wang, et~al.
\newblock Retrieval-augmented generation for large language models: A survey.
\newblock {\em arXiv preprint arXiv:2312.10997}, 2(1):32, 2023.

\bibitem{guu2020retrieval}
Kelvin Guu, Kenton Lee, Zora Tung, Panupong Pasupat, and Mingwei Chang.
\newblock Retrieval augmented language model pre-training.
\newblock In {\em International conference on machine learning}, pages 3929--3938. PMLR, 2020.

\bibitem{hu2022lora}
Edward~J Hu, Yelong Shen, Phillip Wallis, Zeyuan Allen-Zhu, Yuanzhi Li, Shean Wang, Liang Wang, Weizhu Chen, et~al.
\newblock Lora: Low-rank adaptation of large language models.
\newblock {\em Iclr}, 1(2):3, 2022.

\bibitem{huang2024survey}
Yizheng Huang and Jimmy Huang.
\newblock A survey on retrieval-augmented text generation for large language models.
\newblock {\em arXiv preprint arXiv:2404.10981}, 2024.

\bibitem{hui2024uda-wordf1}
Yulong Hui, Yao Lu, and Huanchen Zhang.
\newblock Uda: A benchmark suite for retrieval augmented generation in real-world document analysis.
\newblock {\em Advances in Neural Information Processing Systems}, 37:67200--67217, 2024.

\bibitem{khattab2020colbert}
Omar Khattab and Matei Zaharia.
\newblock Colbert: Efficient and effective passage search via contextualized late interaction over bert.
\newblock In {\em Proceedings of the 43rd International ACM SIGIR conference on research and development in Information Retrieval}, pages 39--48, 2020.

\bibitem{Docmatix}
Hugo Laurençon, Andrés Marafioti, Victor Sanh, and Léo Tronchon.
\newblock Building and better understanding vision-language models: insights and future directions., 2024.

\bibitem{leenv}
Chankyu Lee, Rajarshi Roy, Mengyao Xu, Jonathan Raiman, Mohammad Shoeybi, Bryan Catanzaro, and Wei Ping.
\newblock Nv-embed: Improved techniques for training llms as generalist embedding models.
\newblock In {\em The Thirteenth International Conference on Learning Representations}, 2024.

\bibitem{lewis2020retrieval}
Patrick Lewis, Ethan Perez, Aleksandra Piktus, Fabio Petroni, Vladimir Karpukhin, Naman Goyal, Heinrich K{\"u}ttler, Mike Lewis, Wen-tau Yih, Tim Rockt{\"a}schel, et~al.
\newblock Retrieval-augmented generation for knowledge-intensive nlp tasks.
\newblock {\em Advances in neural information processing systems}, 33:9459--9474, 2020.

\bibitem{ArxivQA}
Lei Li, Yuqi Wang, Runxin Xu, Peiyi Wang, Xiachong Feng, Lingpeng Kong, and Qi Liu.
\newblock Multimodal arxiv: A dataset for improving scientific comprehension of large vision-language models.
\newblock In {\em Proceedings of the 62nd Annual Meeting of the Association for Computational Linguistics (Volume 1: Long Papers)}, pages 14369--14387, 2024.

\bibitem{li2026regionrag}
Yinglu Li, Zhiying Lu, Zhihang Liu, Yiwei Sun, Chuanbin Liu, and Hongtao Xie.
\newblock Regionrag: Region-level retrieval-augmented generation for visual document understanding.
\newblock In {\em Proceedings of the AAAI Conference on Artificial Intelligence}, volume~40, pages 6662--6670, 2026.

\bibitem{lin2004rouge}
Chin-Yew Lin.
\newblock Rouge: A package for automatic evaluation of summaries.
\newblock In {\em Text summarization branches out}, pages 74--81, 2004.

\bibitem{lyu2025crud-textrag2}
Yuanjie Lyu, Zhiyu Li, Simin Niu, Feiyu Xiong, Bo Tang, Wenjin Wang, Hao Wu, Huanyong Liu, Tong Xu, and Enhong Chen.
\newblock Crud-rag: A comprehensive chinese benchmark for retrieval-augmented generation of large language models.
\newblock {\em ACM Transactions on Information Systems}, 43(2):1--32, 2025.

\bibitem{ma2024unifying}
Xueguang Ma, Sheng-Chieh Lin, Minghan Li, Wenhu Chen, and Jimmy Lin.
\newblock Unifying multimodal retrieval via document screenshot embedding.
\newblock In {\em Proceedings of the 2024 Conference on Empirical Methods in Natural Language Processing}, pages 6492--6505, 2024.

\bibitem{mace2025vidore}
Quentin Mac{\'e}, Ant{\'o}nio Loison, and Manuel Faysse.
\newblock Vidore benchmark v2: Raising the bar for visual retrieval.
\newblock {\em arXiv preprint arXiv:2505.17166}, 2025.

\bibitem{ChartQA}
Ahmed Masry, Xuan~Long Do, Jia~Qing Tan, Shafiq Joty, and Enamul Hoque.
\newblock Chartqa: A benchmark for question answering about charts with visual and logical reasoning.
\newblock In {\em Findings of the association for computational linguistics: ACL 2022}, pages 2263--2279, 2022.

\bibitem{InfoVQA}
Minesh Mathew, Viraj Bagal, Rub{\`e}n Tito, Dimosthenis Karatzas, Ernest Valveny, and CV Jawahar.
\newblock Infographicvqa.
\newblock In {\em Proceedings of the IEEE/CVF Winter Conference on Applications of Computer Vision}, pages 1697--1706, 2022.

\bibitem{DocVQA}
Minesh Mathew, Dimosthenis Karatzas, and CV Jawahar.
\newblock Docvqa: A dataset for vqa on document images.
\newblock In {\em Proceedings of the IEEE/CVF winter conference on applications of computer vision}, pages 2200--2209, 2021.

\bibitem{BM25}
Stephen Robertson and Hugo Zaragoza.
\newblock {\em The probabilistic relevance framework: BM25 and beyond}, volume~4.
\newblock Now Publishers Inc, 2009.

\bibitem{shao2024visual-cot}
Hao Shao, Shengju Qian, Han Xiao, Guanglu Song, Zhuofan Zong, Letian Wang, Yu Liu, and Hongsheng Li.
\newblock Visual cot: Advancing multi-modal language models with a comprehensive dataset and benchmark for chain-of-thought reasoning.
\newblock {\em Advances in Neural Information Processing Systems}, 37:8612--8642, 2024.

\bibitem{Tesseract}
Ray Smith.
\newblock An overview of the tesseract ocr engine.
\newblock In {\em Ninth international conference on document analysis and recognition (ICDAR 2007)}, volume~2, pages 629--633. IEEE, 2007.

\bibitem{sorodoc2025garage-textrag4}
Ionut~Teodor Sorodoc, Leonardo~FR Ribeiro, Rexhina Blloshmi, Christopher Davis, and Adri{\`a} de Gispert.
\newblock Garage: A benchmark with grounding annotations for rag evaluation.
\newblock In {\em Findings of the Association for Computational Linguistics: ACL 2025}, pages 17030--17049, 2025.

\bibitem{suri2025visdom}
Manan Suri, Puneet Mathur, Franck Dernoncourt, Kanika Goswami, Ryan~A Rossi, and Dinesh Manocha.
\newblock Visdom: Multi-document qa with visually rich elements using multimodal retrieval-augmented generation.
\newblock In {\em Proceedings of the 2025 Conference of the Nations of the Americas Chapter of the Association for Computational Linguistics: Human Language Technologies (Volume 1: Long Papers)}, pages 6088--6109, 2025.

\bibitem{tanaka2025vdocrag}
Ryota Tanaka, Taichi Iki, Taku Hasegawa, Kyosuke Nishida, Kuniko Saito, and Jun Suzuki.
\newblock Vdocrag: Retrieval-augmented generation over visually-rich documents.
\newblock In {\em Proceedings of the Computer Vision and Pattern Recognition Conference}, pages 24827--24837, 2025.

\bibitem{SlideVQA}
Ryota Tanaka, Kyosuke Nishida, Kosuke Nishida, Taku Hasegawa, Itsumi Saito, and Kuniko Saito.
\newblock Slidevqa: A dataset for document visual question answering on multiple images.
\newblock In {\em Proceedings of the AAAI Conference on Artificial Intelligence}, volume~37, pages 13636--13645, 2023.

\bibitem{ueda2025scan}
Nobuhiro Ueda, Yuyang Dong, Kriszti{\'a}n Boros, Daiki Ito, Takuya Sera, and Masafumi Oyamada.
\newblock Scan: Semantic document layout analysis for textual and visual retrieval-augmented generation.
\newblock In {\em Findings of the Association for Computational Linguistics: EACL 2026}, pages 1618--1637, 2026.

\bibitem{wang2024qwen2vl}
Peng Wang, Shuai Bai, Sinan Tan, Shijie Wang, Zhihao Fan, Jinze Bai, Keqin Chen, Xuejing Liu, Jialin Wang, Wenbin Ge, et~al.
\newblock Qwen2-vl: Enhancing vision-language model's perception of the world at any resolution.
\newblock {\em arXiv preprint arXiv:2409.12191}, 2024.

\bibitem{wang2025vidorag}
Qiuchen Wang, Ruixue Ding, Zehui Chen, Weiqi Wu, Shihang Wang, Pengjun Xie, and Feng Zhao.
\newblock Vidorag: Visual document retrieval-augmented generation via dynamic iterative reasoning agents.
\newblock In {\em Proceedings of the 2025 Conference on Empirical Methods in Natural Language Processing}, pages 9124--9145, 2025.

\bibitem{yang2025qwen3}
An Yang, Anfeng Li, Baosong Yang, Beichen Zhang, Binyuan Hui, Bo Zheng, Bowen Yu, Chang Gao, Chengen Huang, Chenxu Lv, et~al.
\newblock Qwen3 technical report.
\newblock {\em arXiv preprint arXiv:2505.09388}, 2025.

\bibitem{qwen25}
An Yang, Baosong Yang, Beichen Zhang, Binyuan Hui, Bo Zheng, Bowen Yu, Chengyuan Li, Dayiheng Liu, Fei Huang, Haoran Wei, et~al.
\newblock Qwen2.5 technical report, 2025.

\bibitem{yang2024crag-textrag1}
Xiao Yang, Kai Sun, Hao Xin, Yushi Sun, Nikita Bhalla, Xiangsen Chen, Sajal Choudhary, Rongze~D Gui, Ziran~W Jiang, Ziyu Jiang, et~al.
\newblock Crag-comprehensive rag benchmark.
\newblock {\em Advances in Neural Information Processing Systems}, 37:10470--10490, 2024.

\bibitem{yuvisrag}
Shi Yu, Chaoyue Tang, Bokai Xu, Junbo Cui, Junhao Ran, Yukun Yan, Zhenghao Liu, Shuo Wang, Xu Han, Zhiyuan Liu, et~al.
\newblock Visrag: Vision-based retrieval-augmented generation on multi-modality documents.
\newblock In {\em The Thirteenth International Conference on Learning Representations}, 2024.

\bibitem{SigLIP}
Xiaohua Zhai, Basil Mustafa, Alexander Kolesnikov, and Lucas Beyer.
\newblock Sigmoid loss for language image pre-training.
\newblock In {\em Proceedings of the IEEE/CVF international conference on computer vision}, pages 11975--11986, 2023.

\bibitem{zhao2024doclayout}
Zhiyuan Zhao, Hengrui Kang, Bin Wang, and Conghui He.
\newblock Doclayout-yolo: Enhancing document layout analysis through diverse synthetic data and global-to-local adaptive perception.
\newblock {\em arXiv preprint arXiv:2410.12628}, 2024.

\end{thebibliography}
}

\appendix
\section{More Details on Datasets} \label{appendix-A-dataset}

This section provides additional details on the construction of the LFDocQA benchmark used in our experiments, including data sources, annotation procedures, quality control, and implementation details of layout processing.

\subsection{Source Datasets}

\noindent \textbf{Docmatix.}
Docmatix \cite{Docmatix} is a large-scale Document Visual Question Answering (DocVQA) dataset constructed from the PDFA \footnote{https://huggingface.co/datasets/pixparse/pdfa-eng-wds} corpus, which includes 2.1 million raw PDF documents. It comprises 2.4 million document images and 9.5 million QA pairs derived from 1.3 million unique PDFs, covering diverse document types such as reports, forms, and technical documents. Each document page is associated with approximately 4 QA pairs, providing rich supervision for multimodal reasoning.
In our construction, we randomly sample 125k QA pairs along with their corresponding document pages. Among them, 5k samples are used to build the LF-Docmatix benchmark subset, while the remaining 120k samples are used for training data construction.

\noindent \textbf{PaperTab.}
PaperTab is a subset of the UDA (Unstructured Document Analysis) benchmark \cite{hui2024uda-wordf1}, focusing on academic documents with complex layouts. 
It contains 307 research papers in PDF format, with an average of 11 pages and 6.1k words per document. 
The documents are preserved in their original PDF format and typically contain dense tables, figures, and multi-column structures. Compared to Docmatix, PaperTab exhibits more structured and layout-intensive content, making it suitable for evaluating retrieval performance in challenging academic scenarios.
We randomly sample 5k pages from PaperTab to construct the LF-PaperTab benchmark subset.

\subsection{Layout Segmentation and Block Aggregation}
For semantic block aggregation, we introduce three primary thresholds to determine whether adjacent or overlapping bounding boxes should be merged. The specific parameter values and their corresponding descriptions are listed in Table \ref{tab:merge_thresholds}.

\begin{table}[t]
\centering
\caption{Thresholds and parameters used in the bounding box merging algorithm.}
\label{tab:merge_thresholds}
\renewcommand{\arraystretch}{1.2}
\begin{tabular}{p{1.6cm} c p{4.7cm}}
\toprule
\textbf{Parameter} & \textbf{Value} & \textbf{Description} \\
\midrule
$\tau_{y}$ & $40$ pixels & \textbf{Vertical distance threshold.} Merging is considered when the vertical gap between two candidate blocks is less than this value. \\
$\tau_{x}$ & $0.7$ & \textbf{Horizontal alignment threshold.} Merging requires the horizontal Intersection-over-Union (IoU) of the widths of two blocks to be higher than this value. \\
$\tau_{overlap}$ & $0.9$ & \textbf{Physical overlap threshold.} If the intersection area of two blocks accounts for more than this proportion of the smaller block's area, they are considered severely overlapping and are forced to merge. \\
\bottomrule
\end{tabular}
\end{table}

To avoid disrupting the inherent structure of the document, the algorithm restricts merging operations primarily to elements belonging to the same logical semantic group. We define several semantic groups and assign a dominant category to the merged block based on a predefined priority, as illustrated in Table \ref{tab:semantic_groups}.

\begin{table}[t]
\centering
\small
\caption{Semantic grouping of document elements and class preservation priority.}
\label{tab:semantic_groups}
\renewcommand{\arraystretch}{1.1}
\setlength{\tabcolsep}{4pt}
\begin{tabularx}{\columnwidth}{
>{\raggedright\arraybackslash}p{2.05cm}
>{\raggedright\arraybackslash}X
}
\toprule
\textbf{Semantic Group} & \textbf{Original Classes Included} \\
\midrule
Abandon Group & \texttt{abandon} \\
Title Group & \texttt{title} \\
Figure Group & \texttt{figure}, \texttt{figure\_caption} \\
Table Group & \texttt{table}, \texttt{table\_caption}, \texttt{table\_footnote} \\
Text Group & \texttt{plain\_text}, \texttt{isolate\_formula}, \texttt{formula\_caption} \\
\bottomrule
\end{tabularx}
\flushleft \noindent \textit{Note:} Furthermore, the algorithm allows cross-group merging between the Title Group and the Text Group.
\end{table}

\subsection{LLM-based Annotation}
To obtain fine-grained query–block alignment, we adopt a semi-automatic annotation pipeline based on Large Vision-Language Models (VLMs) and Large Language Models (LLMs). 
For LF-Docmatix, given the QA pair and the set of block descriptions, Qwen2.5-72B-Instruct \cite{qwen25} is prompted to select one or multiple blocks that directly support the answer. 
For LF-PaperTab, we follow a complementary strategy. After layout detection, bounding boxes are rendered onto the page to explicitly indicate candidate regions. The annotated page is then fed into Qwen3-VL-32B \cite{bai2025qwen3vl}, which is prompted to generate QA pairs grounded in selected regions.
The detailed prompts used for LF-Docmatix annotation and LF-PaperTab annotation are provided in Figure \ref{prompt:docmatix-annotation} and Figure \ref{prompt:papertab-annotation}, respectively.

\begin{figure}[h]
\centering
\begin{minipage}{\columnwidth}
\begin{tcolorbox}[
    colback=white,
    colframe=black,
    boxrule=0.5pt,
    arc=2pt,
    left=6pt,
    right=6pt,
    top=6pt,
    bottom=6pt
]
\begin{Verbatim}[breaklines=true,breakanywhere=true,fontsize=\scriptsize,breaksymbolleft={},breaksymbolright={}]
You are an expert document understanding assistant. Your core task is to identify the most relevant document regions (bboxes) for a given question-answer pair.
Each bbox represents a meaningful part of the document (text, table, or figure). You need to analyze the question and answer, then select the bboxes that are most relevant to answering the question.
### Detailed Instructions:
- You may use both the textual content of the bboxes and any figure/table descriptions they contain to determine relevance.
- Focus on whether the bbox contains information that directly relates to the question or answer, even if the question itself is not yet suitable for retrieval.
- If no bboxes are relevant to the question, return an empty list.
### Output Format:
You must respond in pure JSON, with the following fields:
{
"relevant_bbox_ids": [list of bbox indices, can be empty if no relevant bboxes],
"reason_for_bboxes": "Explain briefly why these bbox regions are relevant (or why no bboxes are relevant)." 
}
### Output Examples:
**Example 1 (Relevant Bboxes Found):**
{
"relevant_bbox_ids": [1,4],
"reason_for_bboxes": "Bbox 1 identifies the car as 'Tesla Model 3'; bbox 4 lists its weight, which is related to the question 'What is the weight of this car?'."
}
**Example 2 (No Relevant Bboxes):**
{
"relevant_bbox_ids": [],
"reason_for_bboxes": "None of the bboxes contain information related to the question 'What is the melting point of gold?'."
}
Only output the JSON object. No extra text.
\end{Verbatim}
\end{tcolorbox}
\end{minipage}
\caption{Prompt for automatic relevant block annotation in LF-Docmatix.}
\label{prompt:docmatix-annotation}
\end{figure}

\subsection{Query Filtering and Rewriting}
To ensure the queries are suitable for retrieval evaluation, we perform automatic filtering and rewriting using Qwen2.5-7B-Instruct \cite{qwen25}. 
The prompt is shown in Figure \ref{prompt:Query_Filtering_and_Rewriting}.
Specifically, queries are discarded if they:
(1) contain ambiguous references that cannot be grounded to specific regions,
(2) admit multiple valid answers across different documents, or
(3) involve meta-level reasoning unrelated to document content (e.g., counting paragraphs).
For queries that are semantically valid but not explicitly grounded, we rewrite them into structurally localized forms, ensuring that they can be answered using specific document blocks. 
This step improves the alignment between queries and retrieval targets, making the benchmark more suitable for fine-grained evaluation.
After filtering and rewriting, we obtain 87,142 samples for training and 3,487 candidate samples for benchmark construction in LF-Docmatix, along with 4,628 candidate samples for LF-PaperTab.

\subsection{Manual Review Procedure}
\begin{figure*}[t]
  \centering
  \includegraphics[scale=0.8]{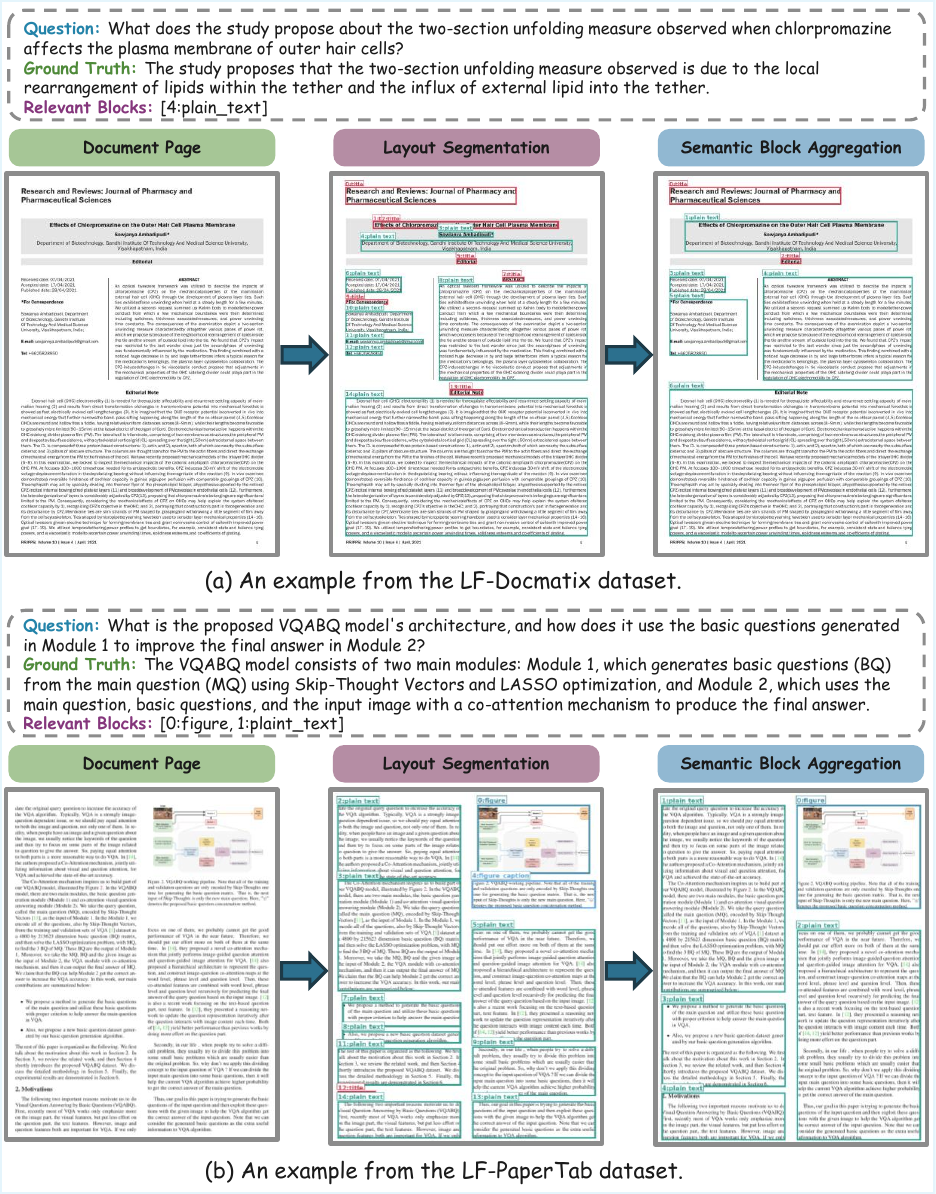}
  \caption{Representative examples from LFDocQA, including document pages, layout segmentation, aggregated semantic blocks, and an annotated QA pair with ground-truth relevant blocks.}
\label{fig:appendix-a-example}
\end{figure*}

To guarantee the reliability and accuracy of the LFDocQA benchmark, we conduct a rigorous manual review procedure involving four professional annotators, divided into two groups (2 annotators per group). 
For each of the two benchmark subsets, we first randomly sample two independent sets of 400 candidate samples, resulting in 800 candidate samples per subset for manual review. Each group of annotators is assigned one set of 400 samples per subset, and all four annotators adhere to the same two evaluation criteria for each sample:
\begin{itemize}
    \item Whether the annotated document blocks correctly support the QA pair;
    \item Whether the query is clear, unambiguous, and grounded in the document content.
\end{itemize}
An sample is retained only if both annotators in the same group agree that it satisfies both criteria. 
The validated samples from the two groups are then merged. 
Finally, we randomly select 500 valid samples from each subset to construct the final LFDocQA benchmark.
As shown in Figure \ref{fig:appendix-a-example}, we provide representative examples for both subsets. 

\begin{figure}[t]
\centering
\begin{minipage}{\columnwidth}
\begin{tcolorbox}[
    colback=white,
    colframe=black,
    boxrule=0.5pt,
    arc=2pt,
    left=6pt,
    right=6pt,
    top=6pt,
    bottom=6pt
]
\begin{Verbatim}[breaklines=true,breakanywhere=true,fontsize=\scriptsize,breaksymbolleft={},breaksymbolright={}]
You are given {len(images)} image snippet(s) retrieved from a document.
Question: {query}
Answer based only on the images.
\end{Verbatim}
\end{tcolorbox}
\end{minipage}
\caption{Prompt for Downstream Answer Generation.}
\label{fig:prompt-for-gen}
\end{figure}

\section{Additional Experiments Details} \label{appendix-B-Experiments}

\subsection{Setup for downstream QA tasks} \label{appendix-B1-Setup}
This section details the prompt templates used for the downstream QA generation task. 
As shown in Figure \ref{fig:prompt-for-gen}, the prompt for the generation model $M_G$ (Qwen2-VL-2B-Instruct \cite{wang2024qwen2vl}) is designed to guide answer generation grounded in the retrieved block images.
For evaluation, we adopt Qwen3-14B \cite{yang2025qwen3} as the automatic judge to score the generated answers on a 0–5 scale (higher scores indicate better quality), with scoring criteria focused on correctness and faithfulness. The judge prompt is carefully designed to eliminate subjective bias and ensure consistent scoring, with detailed prompt in Figure \ref{prompt:llm-judge-prompt}.

\begin{table}[t]
    \centering
    \caption{Generation Performance on LF-Docmatix under different top-$k$ retrieval settings.}
    \small  
    \setlength{\tabcolsep}{1.5pt}
    \begin{tabular}{l  cccc  cccc}
        \toprule
        \multirow{2}{*}{Model} & \multicolumn{4}{c}{Rouge-L} & \multicolumn{4}{c}{ACC (LLM Score)} \\
        & $k$=1 & $k$=3 & $k$=5 & $k$=7 & $k$=1 & $k$=3 & $k$=5 & $k$=7 \\
        \cmidrule(r){1-1} \cmidrule(lr){2-5} \cmidrule(l){6-9} 
        BM25          & \textbf{47.53} & 46.69 & 45.76 & 44.89 & \textbf{2.48} & 2.43 & 2.38 & 2.24 \\
        NV-Embed-V2    & 47.21 & \textbf{47.85} & 46.53 & 45.12 & 2.50 & \textbf{2.57} & 2.49 & 2.44 \\
        BGE-M3         & 47.24 & 47.84 & \textbf{47.96} & 45.64 & 2.38 & \textbf{2.46} & 2.59 & 2.28 \\
        SigLIP         & \textbf{48.76} & 48.26 & 45.60 & 43.79 & \textbf{2.61} & 2.51 & 2.59 & 2.49 \\
        VisRAG         & \textbf{46.05} & 45.79 & 44.36 & 43.89 & \textbf{2.71} & 2.62 & 2.59 & 2.50 \\
        ColPali-v1.1   & 46.54 & \textbf{47.88} & 46.35 & 43.45 & 2.43 & 2.58 & 2.51 & 2.44 \\
        ColQwen2.5     & 46.38 & 47.19 & 44.75 & 45.95 & 2.58 & 2.65 & 2.55 & 2.47 \\
        \textbf{LFRAG (Ours)} & 48.59 & \cellcolor{yellow!20}\textbf{50.78} & 48.68 & 47.13 & 2.68 & \cellcolor{yellow!20}\textbf{2.78} & 2.75 & 2.68 \\
        \bottomrule
    \end{tabular}
    \label{tab:gen-performance-k-Docmatix}
\end{table}

\begin{table}[t]
    \centering
    \caption{Generation Performance  on LF-PaperTab under different top-$k$ retrieval settings.}
    \small  
    \setlength{\tabcolsep}{2pt}
    \begin{tabular}{l cccc cccc}
        \toprule
        \multirow{2}{*}{Model} & \multicolumn{4}{c}{Rouge-L} & \multicolumn{4}{c}{ACC (LLM Score)} \\
        & $k$=1 & $k$=3 & $k$=5 & $k$=7 & $k$=1 & $k$=3 & $k$=5 & $k$=7 \\
        \cmidrule(r){1-1} \cmidrule(lr){2-5} \cmidrule(l){6-9}
        BM25          & 38.57 & \textbf{39.22} & 37.86 & 38.54 & 2.99 & \textbf{3.03} & 3.00 & 3.00 \\
        NV-Embed-V2    & 42.00 & \textbf{42.59} & 41.11 & 40.13 & 3.26 & \textbf{3.31} & 3.22 & 3.22 \\
        BGE-M3         & 41.68 & \textbf{42.88} & 40.75 & 39.50 & 3.26 & \textbf{3.30} & 3.20 & 3.12 \\
        SigLIP         & 35.15 & \textbf{36.57} & 36.77 & 36.72 & 2.64 & 2.74 & 2.79 & \textbf{2.84} \\
        VisRAG         & \textbf{41.40} & 40.76 & 40.16 & 39.74 & \textbf{3.28} & 3.23 & 3.20 & 3.12 \\
        ColPali-v1.1   & 42.11 & \textbf{42.97} & 41.56 & 40.30 & 3.32 & \textbf{3.35} & 3.29 & 3.20 \\
        ColQwen2.5     & \textbf{43.98} & 43.09 & 41.42 & 40.62 & \textbf{3.56} & 3.47 & 3.35 & 3.32 \\
        \textbf{LFRAG (Ours)} & 44.86 & \cellcolor{yellow!20}\textbf{45.37} & 44.13 & 43.58 & 3.57 & \cellcolor{yellow!20}\textbf{3.69} & 3.53 & 3.50 \\
        \bottomrule
    \end{tabular}
    \label{tab:gen-performance-k-papertab}
\end{table}

\subsection{Additional experimental results} \label{appendix-2-results}

\noindent \textbf{Retrieval Performance.}
We present the complete retrieval performance results (nDCG@$k$ and Recall@$k$ for $k=\{1,3,5,10\}$) of all models on the LFDocQA benchmark, supplementing the core nDCG@3 and Recall@3 results (Table \ref{tab:exp1-our-benchmark}) reported in the main text. We report the detailed metrics for both block-level and page-level retrieval across the in-domain subset LF-Docmatix and out-of-domain subset LF-PaperTab.
Tables \ref{tab:exp1-Docmatix-block-level} and \ref{tab:exp1-PaperTab-block-level} show the block-level retrieval performance on LF-Docmatix and LF-PaperTab, respectively. For block-level evaluation, all baselines originally designed for page-level retrieval are adapted by treating each semantically coherent document block as an independent retrieval unit, to align with the fine-grained evaluation setting of LFRAG. Tables \ref{tab:exp1-Docmatix-page-level} and \ref{tab:exp1-PaperTab-page-level} report the page-level retrieval results, where the page similarity score of LFRAG is calculated as the maximum block-level similarity score of all blocks in the target page (Eq.~\ref{page-level-metric} in the main text). Baseline models follow their original page-level retrieval pipelines for consistent comparison.
All metrics are reported as percentage values, and bold values indicate the state-of-the-art (SOTA) performance for the corresponding evaluation setting. Consistent with the main text conclusions, LFRAG outperforms all OCR-based and VLM-based baselines across all $k$ values at both block and page levels, which fully verifies the effectiveness of the layout-oriented fine-grained retrieval strategy in improving the precision and recall of multimodal document retrieval.

\begin{figure*}[t]
  \centering
  \includegraphics[scale=0.8]{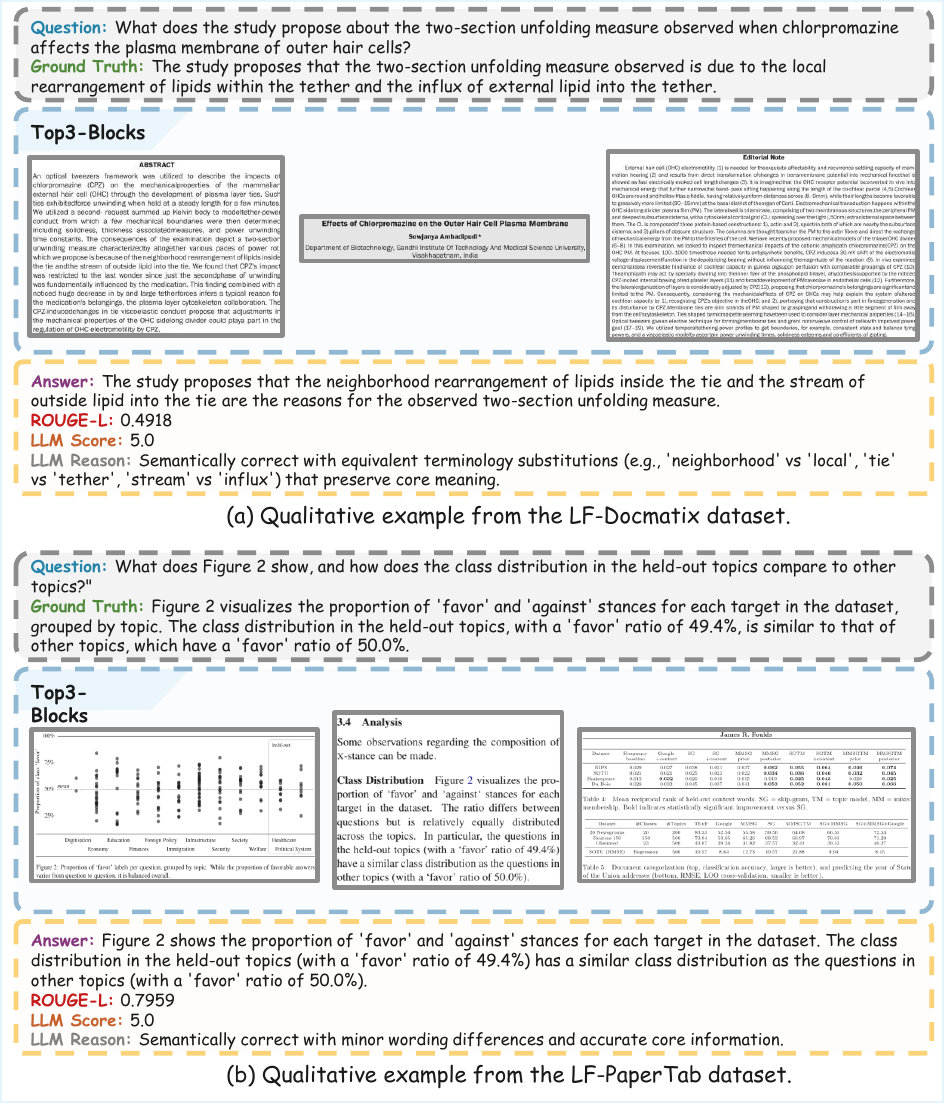}
  \caption{Qualitative examples of LFRAG on LFDocQA. Each case shows the query, top-3 retrieved blocks retrieved by $M_R$, the answer generated by $M_G$, the ground truth answer, and evaluation results.}
\label{fig:appendix-b-example}
\end{figure*}

\noindent \textbf{Generation Performance.}
We report the detailed generation performance of all models on the LFDocQA benchmark under varying top-$k$ retrieved block/page settings (k={1,3,5,7}), with the k=3 results being the core metrics reported in the main text (Table \ref{tab:Generation-LFDocQA}). 
As shown in Table \ref{tab:gen-performance-k-Docmatix} and Table \ref{tab:gen-performance-k-papertab}, we report ROUGE-L and LLM-based accuracy (ACC) on LF-Docmatix and LF-PaperTab, respectively. 
For LFRAG, the metrics correspond to block-level top-$k$ retrieval, while baselines use top-$k$ retrieved page. Bold values indicate the state-of-the-art (SOTA) performance for each $k$ setting, and the yellow highlighted cells mark the peak performance across all models and $k$ values.
Across both subsets, LFRAG achieves SOTA peak performance at $k$=3 for both metrics, realizing an optimal balance between evidence completeness and context redundancy reduction. 
Even at other $k$ values, LFRAG remains competitive and superior to most baselines, whereas all baselines show a general performance decline with increasing $k$, caused by redundant context from page-level retrieval that impairs answer quality and increases hallucination risks.
These results highlight that precise evidence localization, enabled by block-level retrieval and layout-aware representation learning, plays a critical role in improving downstream generation quality. In Figure \ref{fig:appendix-b-example}, we present case examples of LFRAG on LFDocQA.

\begin{table*}[t]
    \centering
    \footnotesize
    \caption{Block-level retrieval performance on LF-Docmatix.}
    \vspace{-1em}
    \begin{tabular}{l l c c c c c c c c}
        \toprule 
        type & Model & nDCG@1 & nDCG@3 & nDCG@5 & nDCG@10 & Recall@1 & Recall@3 & Recall@5 & Recall@10 \\
        \midrule 
        \multirow{3}{*}{\makecell{OCR-based}} & BM25 & 56.00 & 58.26 & 60.61 & 62.14 & 47.31 & 61.31 & 66.98 & 71.40 \\
        & NV-Embed-V2 & 68.20 & 73.74 & 75.73 & 76.84 & 59.01 & 78.48 & 82.98 & 86.21 \\
        & BGE-M3 & 60.20 & 63.46 & 64.37 & 65.50 & 51.61 & 66.71 & 68.76 & 72.06 \\
        \midrule 
        \multirow{5}{*}{\makecell{VLM-based}} & SigLIP & 46.00 & 55.97 & 58.85 & 61.19 & 38.66 & 63.13 & 69.85 & 76.48 \\
        & VisRAG & 73.00 & 81.49 & 82.85 & 84.08 & 64.03 & 87.35 & 90.45 & 93.96 \\
        & ColPali-v1.1 & 75.20 & 78.87 & 80.45 & 82.02 & 66.30 & 82.53 & 86.03 & 90.41 \\
        & ColQwen2.5-v0.2 & 81.16 & 85.06 & 87.08 & 87.73 & 73.66 & 90.23 & 92.48 & 94.31 \\
        & \textbf{LFRAG (Ours)} & \textbf{84.20} & \textbf{90.01} & \textbf{91.29} & \textbf{91.78} & \textbf{74.58} & \textbf{94.08} & \textbf{96.88} & \textbf{98.25} \\
        \bottomrule 
    \end{tabular}
    \label{tab:exp1-Docmatix-block-level}
\end{table*}

\begin{table*}[t]
    \centering
    \footnotesize
    \caption{Block-level retrieval performance on LF-PaperTab.}
    \vspace{-1em}
    \begin{tabular}{l l c c c c c c c c}
        \toprule 
        type & Model & nDCG@1 & nDCG@3 & nDCG@5 & nDCG@10 & Recall@1 & Recall@3 & Recall@5 & Recall@10 \\
        \midrule 
        \multirow{3}{*}{\makecell{OCR-based}} & BM25 & 72.52 & 73.20 & 74.62 & 76.39 & 58.03 & 74.46 & 79.06 & 83.73 \\
        & NV-Embed-V2 & 72.20 & 73.60 & 76.05 & 77.33 & 55.86 & 76.43 & 81.56 & 85.10 \\
        & BGE-M3 & 66.80 & 66.85 & 68.83 & 70.31 & 51.16 & 69.30 & 73.50 & 77.46 \\
        \midrule 
        \multirow{5}{*}{\makecell{VLM-based}} & SigLIP & 27.40 & 30.24 & 33.02 & 36.41 & 19.00 & 33.60 & 39.73 & 49.20 \\
        & VisRAG & 70.60 & 72.54 & 74.63 & 75.82 & 54.36 & 76.33 & 80.66 & 83.96 \\
        & ColPali-v1.1 & 81.75 & 82.60 & 82.91 & 83.81 & 65.10 & 83.73 & 86.16 & 88.60 \\
        & ColQwen2.5-v0.2 & 85.30 & 87.53 & 89.01 & 89.61 & 68.96 & 86.93 & 89.96 & 92.56 \\
        & \textbf{LFRAG (Ours)} & \textbf{90.00} & \textbf{90.84} & \textbf{91.10} & \textbf{91.98} & \textbf{71.73} & \textbf{89.76} & \textbf{92.30} & \textbf{94.73} \\
        \bottomrule 
    \end{tabular}   
    \label{tab:exp1-PaperTab-block-level}
\end{table*}

\begin{table*}[t]
    \centering
    \footnotesize
    \caption{Page-level retrieval performance on LF-Docmatix.}
    \vspace{-1em}
    \begin{tabular}{l l c c c c c c c c}
        \toprule 
        type & Model & nDCG@1 & nDCG@3 & nDCG@5 & nDCG@10 & Recall@1 & Recall@3 & Recall@5 & Recall@10 \\
        \midrule 
        \multirow{3}{*}{\makecell{OCR-based}} & BM25 & 70.20 & 76.19 & 77.26 & 78.32 & 70.20 & 80.40 & 83.00 & 86.20 \\
        & NV-Embed-V2 & 79.00 & 84.72 & 85.72 & 86.31 & 79.00 & 88.60 & 91.00 & 92.80 \\
        & BGE-M3 & 72.40 & 77.80 & 79.79 & 81.27 & 72.40 & 82.00 & 86.80 & 91.40 \\
        \midrule 
        \multirow{5}{*}{\makecell{VLM-based}} & SigLIP & 72.80 & 79.55 & 81.71 & 82.96 & 72.80 & 83.80 & 89.00 & 92.80 \\
        & VisRAG & 87.60 & 91.78 & 92.44 & 92.84 & 87.60 & 94.60 & 96.20 & 97.40 \\
        & ColPali-v1.1 & 83.00 & 87.81 & 88.55 & 89.24 & 83.00 & 91.00 & 92.80 & 95.00 \\
        & ColQwen2.5-v0.2 & 87.50 & 92.65 & 93.98 & 94.90 & 88.00 & 95.00 & 96.80 & 97.80 \\
        & \textbf{LFRAG (Ours)} & \textbf{89.60} & \textbf{94.44} & \textbf{94.86} & \textbf{95.12} & \textbf{89.60} & \textbf{97.40} & \textbf{98.40} & \textbf{99.20} \\
        \bottomrule 
    \end{tabular}
    \label{tab:exp1-Docmatix-page-level}
\end{table*}

\begin{table*}[t]
    \centering
    \footnotesize
    \caption{Page-level retrieval performance on LF-PaperTab.}
    \vspace{-1em}
    \begin{tabular}{l l c c c c c c c c}
        \toprule 
        type & Model & nDCG@1 & nDCG@3 & nDCG@5 & nDCG@10 & Recall@1 & Recall@3 & Recall@5 & Recall@10 \\
        \midrule 
        \multirow{3}{*}{\makecell{OCR-based}} & BM25 & 78.40 & 84.49 & 84.96 & 85.81 & 78.40 & 88.80 & 90.00 & 92.60 \\
        & NV-Embed-V2 & 73.20 & 78.23 & 79.23 & 80.63 & 73.20 & 81.80 & 84.20 & 88.40 \\
        & BGE-M3 & 58.60 & 64.73 & 66.31 & 68.27 & 58.60 & 69.20 & 73.00 & 79.00 \\
        \midrule 
        \multirow{5}{*}{\makecell{VLM-based}} & SigLIP & 39.80 & 48.13 & 50.68 & 53.33 & 39.80 & 54.00 & 60.20 & 68.40 \\
        & VisRAG & 77.80 & 82.68 & 84.11 & 84.85 & 77.80 & 86.00 & 89.40 & 91.80 \\
        & ColPali-v1.1 & 82.60 & 86.95 & 87.88 & 88.65 & 82.60 & 90.00 & 92.20 & 94.60 \\
        & ColQwen2.5-v0.2 & 90.60 & 91.88 & 92.52 & 93.71 & 90.60 & 92.80 & 95.40 & 96.50 \\
        & \textbf{LFRAG (Ours)} & \textbf{92.60} & \textbf{94.54} & \textbf{94.86} & \textbf{95.45} & \textbf{92.60} & \textbf{95.80} & \textbf{96.60} & \textbf{98.40} \\
        \bottomrule 
    \end{tabular}
    \label{tab:exp1-PaperTab-page-level}
\end{table*}

\begin{figure*}[h]
\centering
\begin{tcolorbox}[
    colback=white,
    colframe=black,
    boxrule=0.5pt,
    arc=2pt,
    left=6pt,
    right=6pt,
    top=2pt,
    bottom=2pt,
]
\begin{Verbatim}[breaklines=true,breakanywhere=true,fontsize=\scriptsize,breaksymbolleft={},breaksymbolright={}]
You are an expert in generating training data for multimodal Retrieval-Augmented Generation (RAG) models.
Carefully analyze the provided document page. The image has been pre-annotated: different content types are marked with red bounding boxes and labeled in the format 'classname_ID' (e.g., 'title_4', 'plain text_1'). These annotations are for your reference only, to help you identify the evidence.
Your task is to generate 2-3 high-quality Question-Answering (QA) pairs based on the content and layout of the image.

Please adhere to the following rules:
1.  **Prioritize Relational Questions**: Strive to create questions that require synthesizing information from **multiple (two or three or four) bounding boxes** to answer. These are high-value for RAG training.
2.  **Allow High-Quality Single-Box Questions**: You may also include questions that can be answered from a single box, but only if the question requires summarization, inference, or understanding the text's purpose (not just simple text extraction).
3.  **Leverage Layout and Semantic Relationships**: Generate questions that require reasoning about the relationship between different content types based on their visual layout and semantic meaning. For example:
    *   Ask how a figure is explained by its nearby caption.
    *   Ask what the text under a specific title (e.g., "Introduction") describes.
    *   Ask to summarize the content of a table based on its accompanying caption.
4.  **Identify Evidence IDs**: Each box is labeled in the format `classname_ID`. Your output JSON must identify the relevant ID numbers (the integers) required to answer the question in the `relevant_bbox_ids` field.
5.  **Do Not Expose Annotation Labels**: The `question` and `answer` strings you generate must be natural and sound as if a human is asking about the document. They must not contain the `classname_ID` labels (e.g., 'title_4', 'plain text_1'). These labels are only for your internal use to populate the `relevant_bbox_ids` list.
6.  **Avoid General/Retrieval-Unfriendly Questions**: Do not ask overly broad questions about the document's structure or simple facts that are not suitable for a retrieval task. For example, avoid questions like "How many tables are in this paper?" or "What is the title of section 2?". Focus on the specific content and its meaning.
7.  **JSON Output Format**: Your response must be a strictly formatted JSON object. Do not include any explanatory text before or after the JSON.
8.  **JSON Structure**: The JSON object must contain a single key "QAs", which holds a list of objects. Each object in the list must contain three keys: `question` (string), `answer` (string), and `relevant_bbox_ids` (a list of integers).

Example of the required output format:
```json
{
    "QAs": [
        {
            "question": "What are the affiliation and email address of the author Biao Liu?",
            "answer": "His affiliation is Tsinghua University and his email is liubiao2638@gmail.com.",
            "relevant_bbox_ids": [10, 8]
        },
        {
            "question": "Explain the objective function for GE-FL shown in formula (2).",
            "answer": "xxxxxx",
            "relevant_bbox_ids": [14, 3, 2]
        },
    ]
}
```
\end{Verbatim}
\end{tcolorbox}
\caption{Prompt for grounded QA generation in LF-PaperTab.}
\label{prompt:papertab-annotation}
\end{figure*}

\begin{figure}[t]
\centering
\begin{minipage}{\columnwidth}
\begin{tcolorbox}[
    colback=white,
    colframe=black,
    boxrule=0.5pt,
    arc=2pt,
    left=6pt,
    right=6pt,
    top=6pt,
    bottom=6pt
]
\begin{Verbatim}[breaklines=true,breakanywhere=true,fontsize=\scriptsize,breaksymbolleft={},breaksymbolright={}]
You are an expert document retrieval and question refinement assistant. You will be given a question-answer pair, then judge whether the question can be rewritten into a valid retrieval question or should be discarded. Your core task is to convert a specific-document QA pair into a document retrieval QA pair.
### Detailed Instructions:
- **Step 1: Evaluate and refine the question**
Analyze the question and determine if it qualifies as a valid retrieval question, i.e., whether it can uniquely point to specific documents in a large collection based solely on its own content.
A valid retrieval question must be self-contained and document-agnostic: it should include specific, unambiguous information (e.g., unique names, titles, events, identifiers) that allows identifying relevant documents without relying on implicit references to an unspecified "current document."
1. If the question is already self-contained and document-agnostic, keep it unchanged.
2. If the question is ambiguous or context-dependent (e.g., uses "the document", "the product"), use the relevant bboxes to find specific information (e.g., unique names, identifiers) and rewrite it into a standalone, document-independent question.
*When rewriting, keep the new question as concise as possible while preserving the original meaning.* 
3. If it cannot be rewritten into a clear standalone form, discard it (set question to null).
- **Step 2: Output your final judgment in pure JSON.**
### Output Format:
You must respond in pure JSON, with the following fields:
{
"question": "rewritten or original question text, or null if discarded",
"reason_for_question": "Explain briefly why you kept, rewrote, or discarded the question (reference relevant bboxes if used for rewriting)." 
}
### Output Examples:
**Example 1 (Retained Only Because Fully Standalone):**
{
"question": "What is the melting point of gold?",
"reason_for_question": "No ties to the current document; references a universal entity ('gold') fully retrievable; no rewriting needed."
}
**Example 2 (Rewritten Using Relevant Bboxes):**
{
"question": "What is the weight of the Tesla Model 3?",
"reason_for_question": "Original question 'What is the weight of this car?' contained 'this car' (tied to the document); rewritten using 'Tesla Model 3' (specific entity from relevant bbox 1)." 
}
**Example 3 (Discarded Unfixable Document Dependence):**
{
"question": null,
"reason_for_question": "Original question 'What is the first section about?' relies on 'the first section' (specific to the current document), and no relevant bboxes provide specific information to rewrite it into a standalone question."
}
Only output the JSON object. No extra text.
\end{Verbatim}
\end{tcolorbox}
\end{minipage}
\caption{Prompt for query filtering and rewriting.}
\label{prompt:Query_Filtering_and_Rewriting}
\end{figure}

\begin{figure}[t]
\centering
\begin{minipage}{\columnwidth}
\begin{tcolorbox}[
    colback=white,
    colframe=black,
    boxrule=0.5pt,
    arc=2pt,
    left=6pt,
    right=6pt,
    top=6pt,
    bottom=6pt
]
\begin{Verbatim}[breaklines=true,breakanywhere=true,fontsize=\scriptsize,breaksymbolleft={},breaksymbolright={}]
You are a QA evaluator. Compare the [Reference Answer] and the [Model Answer], then assign an integer score from 0 to 5.
Scoring rubric:
- 5: Semantically correct and covers the core information; different wording/format and minor irrelevant additions are acceptable.
- 4: Mostly correct with only minor omissions or minor inaccuracies that do not affect the main conclusion.
- 3: Partially correct; includes some key information but has clear missing parts or local errors.
- 2: Only limited relevant content; most key facts are missing or confused.
- 1: Mostly incorrect; only weak relevance.
- 0: Completely incorrect, off-topic, contradictory to the reference answer, or no valid answer.
Additional rules:
1) For numbers, dates, and proper nouns: if the main conclusion is correct but formatting differs slightly (e.g., 09/01 vs September 1), be lenient.
2) For long answers: if the core conclusion is correct but contains a small amount of noise, do not over-penalize.
3) If the reference answer indicates "not mentioned/cannot be determined", then fabricated specific facts should receive a low score (0-2).

Return strict JSON only. Do not output anything else:
{{"score": <integer 0-5>, "reason": "one short English reason"}}

[Reference Answer]
{ground_truth}
[Model Answer]
{prediction}
\end{Verbatim}
\end{tcolorbox}
\end{minipage}
\caption{Prompt for LLM-based QA Answer Evaluation.}
\label{prompt:llm-judge-prompt}
\end{figure}

\end{document}